\documentclass[manuscript,screen]{acmart}

\acmJournal{TQC}
\acmDOI{}
\acmISBN{}

\usepackage{mathtools}
\usepackage{algorithm}
\usepackage{algorithmic}
\usepackage{array}
\usepackage{placeins}
\graphicspath{{./}}
\newcommand{\ket}[1]{\lvert #1 \rangle}

\newcommand{\braket}[2]{\langle #1 \mid #2 \rangle}
\newcommand{\expect}[1]{\langle #1 \rangle}
\newcommand{\qpd}{\mathrm{QPD}}

\newcommand{\rmultichi}{R_{\mathrm{multi},\chi}}
\newcommand{\localinf}{L_{\chi}}
\newcommand{\epcost}{C_{\mathrm{EP}}}
\newcommand{\gatecutscore}{C_{\mathrm{gate}}}
\newcommand{\qpdoverhead}{\Omega_{\mathrm{QPD}}}
\newcommand{\hellingerfid}{F_{\mathrm{H}}}
\newcommand{\hellingerinf}{I_{\mathrm{H}}}

\definecolor{wfblue}{RGB}{31,119,180}
\definecolor{wforange}{RGB}{230,109,0}
\definecolor{wfline}{RGB}{65,72,78}
\definecolor{wfgraph}{RGB}{48,82,110}
\definecolor{wftext}{RGB}{25,30,35}

\begin{document}

\title{Entanglement-Capability-Aware Finite-Bond Circuit Cutting for Distributed Quantum Computing}

\author{Weidong Liu}
\email{weidong2024@126.com}
\affiliation{%
  \institution{Zhengzhou University, Zhengzhou}
  \city{Zhengzhou}
  \country{China}}

\author{Qingqing Jiang}
\email{anzhi2710@gmail.com}
\affiliation{%
  \institution{Zhengzhou University, Zhengzhou}
  \city{Zhengzhou}
  \country{China}}

\author{Yufu Liu}
\email{2732756895@qq.com}
\affiliation{%
  \institution{Zhengzhou University, Zhengzhou}
  \city{Zhengzhou}
  \country{China}}

\author{Ruiqing He}
\email{njheruiqing@hotmail.com}
\affiliation{%
  \institution{Nanjing Institute of Technology, Nanjing}
  \city{Nanjing}
  \country{China}}

\author{Jiandong Shang}
\email{sjd@zzu.edu.cn}
\affiliation{%
  \institution{Zhengzhou University, Zhengzhou}
  \city{Zhengzhou}
  \country{China}}

\author{Hengliang Guo}
\email{guohengliang@zzu.edu.cn}
\affiliation{%
  \institution{Zhengzhou University, Zhengzhou}
  \city{Zhengzhou}
  \country{China}}

\author{Qiang Chen}
\email{qiangchen@zzu.edu.cn}
\affiliation{%
  \institution{Zhengzhou University, Zhengzhou}
  \city{Zhengzhou}
  \country{China}}

\authorsaddresses{This work is supported by the National Key Research and Development Program of China (2024YFB4504103).\\ Authors' Contact Information: Weidong Liu, National Supercomputing Center in Zhengzhou and School of Computer and Artificial Intelligence, Zhengzhou University, Zhengzhou, China; e-mail: weidong2024@126.com; Qingqing Jiang, National Supercomputing Center in Zhengzhou and School of Computer and Artificial Intelligence, Zhengzhou University, Zhengzhou, China; e-mail: anzhi2710@gmail.com; Yufu Liu, National Supercomputing Center in Zhengzhou and School of Computer and Artificial Intelligence, Zhengzhou University, Zhengzhou, China; e-mail: 2732756895@qq.com; Ruiqing He, School of Communication and Artificial Intelligence and the School of Integrated Circuits, Nanjing Institute of Technology, Nanjing, China; e-mail: njheruiqing@hotmail.com; Jiandong Shang, National Supercomputing Center in Zhengzhou, Zhengzhou University, Zhengzhou, China; e-mail: sjd@zzu.edu.cn; Hengliang Guo, National Supercomputing Center in Zhengzhou, Zhengzhou University, Zhengzhou, China; e-mail: guohengliang@zzu.edu.cn; Qiang Chen (corresponding author), National Supercomputing Center in Zhengzhou, Zhengzhou University, Zhengzhou, China; e-mail: qiangchen@zzu.edu.cn}

\begin{abstract}
Distributed quantum computing distributes workloads that exceed a single processor's capacity across resource-constrained components. Circuit cutting enables this decomposition, but its effectiveness depends on partition choice. We present SACE, a Simulation-Aware Circuit Partitioning framework with Entanglement-Capability-Aware Screening. EP-Cost screens partitions using a parameter-dependent entanglement-capability proxy and native implementation cost. Under reconstruction constraints, finite-bond refinement selects the candidate that minimizes the worst-subcircuit discrepancy between normalized matrix product state (MPS) simulations at bond dimensions $\chi$ and $2\chi$. Across 14 exhaustively enumerated small-scale candidate spaces, EP-Cost-guided genetic algorithm selections attain co-optimal native quasiprobability decomposition (QPD) overhead, with a median within-circuit Spearman correlation of 0.9965 between EP-Cost and QPD overhead. At matched cut counts, EP-Cost reduces QPD overhead by $25.20\times$ and $16.89\times$ for 8- and 9-qubit quantum Fourier transform (QFT) circuits, respectively, and improves sampled reconstruction. On 12-qubit approximate QFT (AQFT), deterministic reconstruction reduces Pauli-observable mean absolute error in all six low-bond conditions and total-variation distance in five. With five cuts and 7,776 QPD terms, SACE is best on all four reconstruction metrics in four of six conditions. At 24 qubits, SACE improves exact-local QFT accuracy in all 30 instance-bond-dimension conditions against each EP-Cost-only, Kernighan-Lin, and spectral baseline. The 50-qubit QFT comparisons remain favorable as reference convergence improves. Weighted quantum approximate optimization algorithm (QAOA) circuits provide complementary, more bond-sensitive evidence. Refinement benefits persist across three search methods; larger candidate panels improve the optimized proxy without guaranteeing lower exact-local error. These results support reconstruction-aware boundary screening followed by finite-bond refinement.
\end{abstract}



\maketitle

\section{Introduction}
Quantum computing is progressing from the noisy intermediate-scale quantum (NISQ) regime toward fault-tolerant systems, while individual processors remain limited in capacity and reliability~\cite{Preskill2018nisq}. Quantum applications also require classical computing for compilation, control, simulation, workflow orchestration, and result processing. These needs motivate the integration of quantum processing units (QPUs) with high-performance computing (HPC) infrastructure~\cite{Beck2024hpcecosystems,Du2025qccp,Shehata2026convergence}. QPUs supply specialized quantum capabilities, while HPC resources coordinate data processing, resource management, and iterative quantum--classical execution.

Hybrid quantum--HPC infrastructure is being developed in several countries. In Europe, the EuroHPC Joint Undertaking is integrating quantum systems based on several hardware technologies with existing supercomputers~\cite{EuroHPC2026QuantumAccess}. In Japan, RIKEN has begun collaborative operation between Fugaku and IBM Quantum System Two and is developing a broader quantum--HPC platform~\cite{RIKEN2025FugakuQuantum}. In the United States, Oak Ridge National Laboratory is developing systems that connect quantum processors based on different hardware technologies with leadership-class HPC resources~\cite{ORNL2025QuantumHPC}.

Our broader work follows this direction. Sugon 8000 (Dengfeng) integrates scientific computing and artificial intelligence (AI) workloads~\cite{Sugon2026Dengfeng}. Ongoing work aims to tightly couple this large-scale domestic infrastructure with quantum computers based on multiple hardware technologies, supporting cooperation among quantum, HPC, and AI resources. Circuits that exceed a single QPU's capacity can be divided into smaller tasks, with classical resources coordinating their execution. Distributed quantum computing provides a basis for this heterogeneous workflow~\cite{Cuomo2020ecosystem,Caleffi2024survey}. Circuit cutting enables the decomposition by dividing a circuit into smaller subcircuits and reconstructing the target computation through classical post-processing~\cite{Peng2020smallqc,Mitarai2021virtualgate,Piveteau2024knitting}. This paper addresses the compiler-level partition-selection step before subcircuit execution and classical reconstruction.

Circuit cutting transfers computational work to classical reconstruction. Its cost depends on the boundary operations, since quasiprobability decomposition (QPD) branching and coefficient growth vary with gate type and parameters. Quantum--HPC studies treat execution and reconstruction as coupled workflow stages, and multi-node implementations using central processing units and graphics processing units show that reconstruction can require substantial classical resources~\cite{Du2025qccp,Garcia2024multinode}. The partition also determines local subcircuit workloads, so selection must account for both reconstruction and local execution or simulation resources.

Existing distributed-compilation methods commonly optimize communication volume, teleportation count, or weighted cut size through graph or hypergraph models~\cite{Daei2020optimized,Baker2020timesliced,Burt2024generalised,Burt2026multilevel}. Circuit-cutting methods also use cut count and gate type to control QPD overhead~\cite{Mitarai2021overhead,Piveteau2022reduced,Piveteau2024knitting}. For parameterized circuits, equal cut counts can incur different reconstruction costs because the cut gates have different parameters. Graph objectives that weight these interactions equally cannot capture this difference.

Partition choice also affects local simulation accuracy. A matrix product state (MPS) simulation depends on the correlations retained within each subcircuit and the available bond dimension~\cite{Vidal2003mps,Schollwock2011mps}. Inexpensive cuts may leave subcircuits that are difficult to approximate, while improved local representability may require more costly reconstruction. We therefore evaluate boundary cost and finite-bond behavior in separate stages.

We propose SACE, a Simulation-Aware Circuit Partitioning framework with Entanglement-Capability-Aware Screening. Its first stage uses EP-Cost to rank candidate boundaries according to a gate-level entanglement-capability proxy and native implementation cost. The second stage evaluates the retained candidates by simulating their isolated local subcircuits at bond dimensions $\chi$ and $2\chi$. SACE selects the partition with the smallest worst-subcircuit discrepancy, using EP-Cost and a canonical partition identifier to resolve ties.

Constraints on cut count, QPD term count, or QPD overhead restrict the candidate set before finite-bond selection, without changing the ranking rule. The execution environment supplies these resource limits. We evaluate SACE as a compiler- and simulation-level component within a quantum--HPC workflow; a complete runtime and scheduling system are outside the scope of this study.

The contributions are as follows:
\begin{enumerate}
  \item[(1)] We introduce a two-stage partitioning framework that screens candidates by boundary cost and refines them by finite-bond MPS behavior, with reconstruction budgets imposed as feasibility constraints.
  \item[(2)] We define EP-Cost from gate parameters and native implementation costs, and use the worst-subcircuit discrepancy between $\chi$- and $2\chi$-bond MPS simulations to select among the retained candidates.
  \item[(3)] We evaluate the framework through a layered validation strategy: exhaustive candidate-space and sampled-QPD studies at small scale, complete deterministic-QPD reconstruction on a 12-qubit approximate quantum Fourier transform (AQFT) circuit, exact-local finite-bond validation at 24 qubits, and higher-resource MPS consistency analysis at 50 qubits. We further compare genetic algorithm, simulated annealing, and discrete particle swarm optimization searches under matched search budgets.
\end{enumerate}

The evaluations address different levels of the workflow. The 12-qubit AQFT study performs complete deterministic-QPD reconstruction and evaluates reconstructed distributions and observables. At 24 qubits, exact isolated-subcircuit states provide local references for finite-bond accuracy. At 50 qubits, validation uses higher-resource MPS references and explicitly reports their convergence. Quantum Fourier transform (QFT) circuits provide the primary scale-up evidence as reference convergence improves with $\chi$; reference-limited weighted quantum approximate optimization algorithm (QAOA) results are reported in the appendix. The 24- and 50-qubit studies evaluate local finite-bond behavior, not reconstructed global-circuit fidelity.
\section{Quantum Circuit Cutting and Reconstruction}

\subsection{Circuit-cutting workflow}
Consider a quantum circuit whose width or execution requirements exceed the capacity of a single local quantum resource. A partition assigns the logical qubits to capacity-constrained blocks. Operations contained within one block remain local, whereas two-qubit operations that cross block boundaries must be handled by a distributed execution mechanism such as circuit cutting. After cutting, the resulting subcircuits can be executed independently on available quantum processors, while classical resources coordinate the reconstruction of the target quantity~\cite{Cuomo2020ecosystem,Caleffi2024survey,Peng2020smallqc,Tang2021cutqc}.

Distributed-compilation methods use graph or hypergraph models to reduce communication, teleportation, or entanglement-distribution costs under hardware-capacity constraints~\cite{Daei2020optimized,Baker2020timesliced,Burt2024generalised,Burt2026multilevel}. In circuit cutting, the partition also determines classical reconstruction work and local subcircuit structure. Selection must therefore account for both the reconstruction cost of boundary operations and the tractability of the resulting subcircuits under available execution or simulation resources.

\begin{figure}[!htbp]
  \centering
  \includegraphics[width=\linewidth,keepaspectratio]{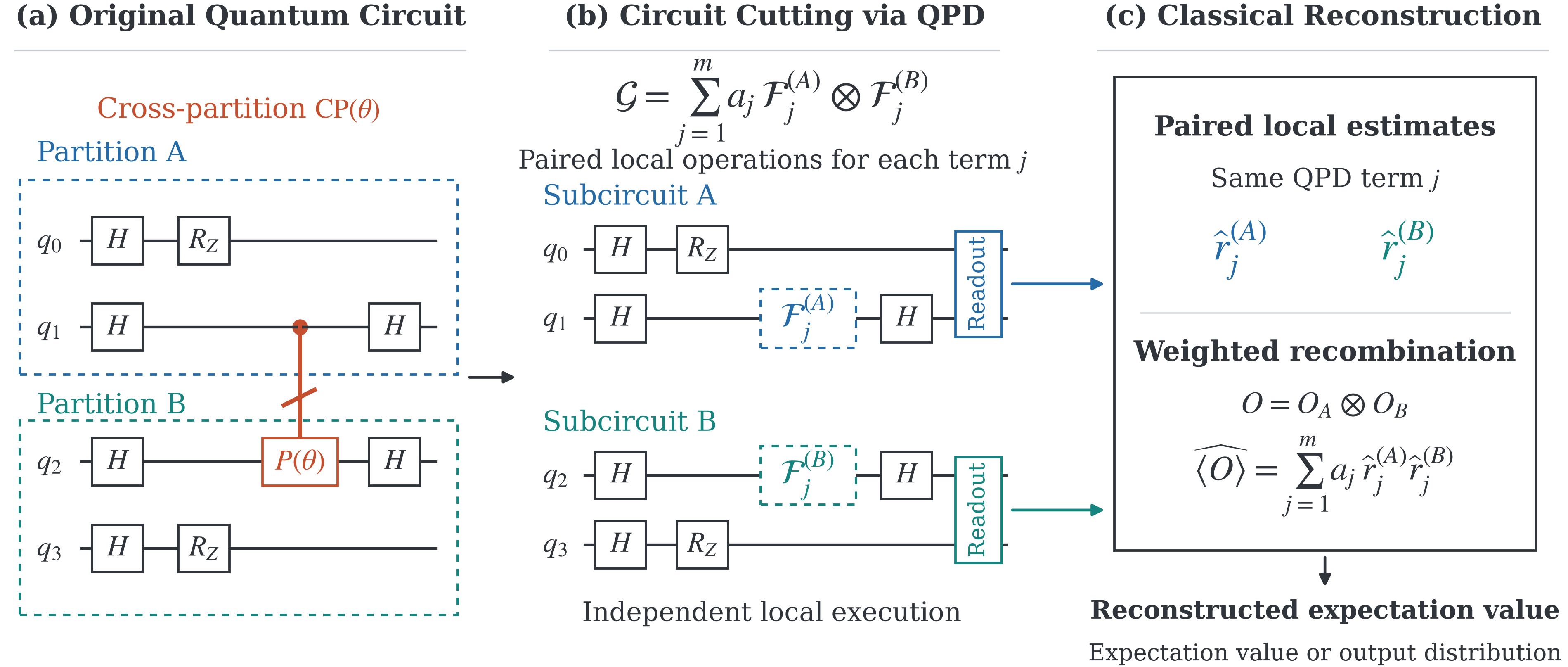}
  \caption{Schematic illustration of quantum circuit cutting and classical reconstruction. (a) A controlled-phase gate crosses two qubit partitions. (b) Quasiprobability decomposition expresses the cross-partition channel as weighted pairs of local operations executed in separate subcircuits. (c) The paired local estimates are combined through classical post-processing to reconstruct the expectation value of a product observable.}
  \Description[Quantum circuit cutting and classical reconstruction]{Three panels show a four-qubit quantum circuit split into two partitions, with a cross-partition controlled-phase gate marked for cutting; QPD replaces the cut gate with corresponding local operations in two independent subcircuits; paired local readouts are combined using QPD coefficients to reconstruct a product-observable expectation value.}
  \label{fig:circuit-cutting-reconstruction}
\end{figure}

Figure~\ref{fig:circuit-cutting-reconstruction} illustrates the circuit-cutting and classical reconstruction workflow.

\subsection{Reconstruction cost of a circuit partition}
For a partition $P$, let $\delta(P)$ denote the set of two-qubit gates that cross partition boundaries. For a cross-partition quantum channel $\mathcal{G}$, a QPD represents the channel as
\begin{equation}
  \mathcal{G} = \sum_{j=1}^{m} a_j\,\mathcal{F}^{(A)}_j \otimes \mathcal{F}^{(B)}_j,
  \label{eq:qpd}
\end{equation}
where the coefficients $a_j$ may be negative and the local channels act independently on the two sides of the cut. Sampling terms in Eq.~\eqref{eq:qpd} gives an unbiased estimator of the target observable, with variance controlled by
\begin{equation}
  \gamma(\mathcal{G}) = \sum_{j=1}^{m} |a_j|.
\end{equation}
For multiple independent cuts, the aggregate factor is multiplicative,
\begin{equation}
  \Gamma(P)=\prod_{g\in\delta(P)}\gamma_g,
\end{equation}
and the sampling overhead scales quadratically with $\Gamma(P)$ for fixed estimation error~\cite{Mitarai2021overhead,Piveteau2022reduced}. We use $\Omega_{\mathrm{QPD}}$ in later sections for the corresponding native QPD overhead.

This multiplicative structure makes the partition boundary a direct source of reconstruction cost. Early resource-trading and fragment-based approaches showed how smaller quantum processors can be combined with classical post-processing~\cite{Bravyi2016trading,Peng2020smallqc}. Circuit-cutting systems then developed this idea into practical fragment execution and reconstruction workflows~\cite{Ayral2020divide,Tang2021cutqc}. Subsequent work has optimized the placement and type of gate and wire cuts~\cite{Brandhofer2024optimal,Chen2023golden,Basu2024fragqc,Nakamura2025beyond}. QPD-based analyses have also characterized or reduced the overhead of nonlocal gates~\cite{Mitarai2021overhead,Piveteau2022reduced,Piveteau2024knitting,Gentinetta2024overhead} and wire or multiple-cut constructions~\cite{Harada2024wire,Schmitt2025multiple,Brenner2025wire,Jing2025entanglementcost}. Optimized combinations of gate cuts, wire cuts, ancillas, and classical communication can further reduce this cost~\cite{Brandhofer2024optimal,Piveteau2024knitting}, while randomized-measurement methods provide a different trade-off between cut structure and sample complexity~\cite{Lowe2023fast}.

These results also show why cut count alone is incomplete. Two partitions may satisfy
\begin{equation}
  |\delta(P_1)|=|\delta(P_2)|
\end{equation}
while incurring different reconstruction costs because the crossed operations differ in gate type, parameter values, decomposition, or communication requirements. Partition selection therefore requires more information than the number of boundary gates.

\subsection{Local tractability under finite bond dimension}
The same partition also determines the local circuits that remain after cross-partition operations are removed or decomposed. Even when two partitions have similar reconstruction cost, their local subcircuits can retain different correlation structures and can therefore differ in the resources required for accurate simulation or execution-assisted validation.

An MPS represents an $n$-qubit state by local tensors connected through bond indices. The bond dimension $\chi$ limits the correlations that can be represented and strongly affects simulation cost~\cite{Vidal2003mps,Schollwock2011mps,Cirac2021mps}. Circuits of similar width can consequently have different finite-bond accuracy and resource requirements~\cite{Barratt2021parallel,Berezutskii2025tensor,Leonteva2026emulators}. Comparing normalized local simulations at increasing bond dimensions provides a practical convergence diagnostic for a retained subcircuit.

Partition selection must balance reconstruction cost against local finite-bond tractability: inexpensive boundaries can leave difficult local simulations, while more tractable subcircuits may require costly reconstruction.

\subsection{Graph-based partition model and problem statement}
Let a quantum circuit $C$ act on logical qubits $V=\{q_1,\ldots,q_n\}$. We represent the circuit by a weighted interaction graph
\begin{equation}
  G_C=(V,E,w,t),
\end{equation}
where an edge $e=(u,v)$ represents one or more two-qubit gate instances between qubits $u$ and $v$, $w_e$ denotes the corresponding interaction weight, and $t_e$ retains the gate-type and parameter information needed by downstream cost models. Repeated interactions may be represented as separate edges or aggregated, provided that this metadata is preserved. Single-qubit gates do not contribute interaction edges but remain part of the corresponding local subcircuits.

A $K$-way partition is $P=\{V_1,\ldots,V_K\}$ with disjoint blocks whose union is $V$. Each block satisfies a capacity constraint $|V_k|\leq Q_k$, where $Q_k$ denotes the available qubits or the chosen local simulation width. The formulation permits general $K$, while the SACE implementation and all evaluations in this work use capacity-constrained bipartitions ($K=2$). Under this model, the boundary set $\delta(P)$ introduced above contains the two-qubit gates whose endpoints lie in different blocks.

Given a circuit $C$, local capacity limits $\{Q_k\}$, a finite local-resource budget, and optional reconstruction constraints, we seek a feasible partition with acceptable reconstruction cost and tractable local subcircuits. Section~3 addresses these requirements through boundary screening followed by finite-bond refinement.

\section{SACE: Entanglement-Capability-Aware Finite-Bond Resource-Guided Partitioning}

\subsection{Entanglement-capability-aware boundary score}
Let $\mathcal{G}_2(C)$ denote the set of two-qubit gate instances in circuit $C$. For each gate $g$, let $\eta_g\geq0$ denote the parameter-dependent gate-level entanglement-capability proxy used by EP-Cost, and let $\kappa_g\geq0$ denote its cost under the fixed native two-qubit-gate model. For the controlled-phase (CP), $R_{ZZ}$, controlled-NOT (CX), controlled-$Z$ (CZ), and SWAP gates used in this work,
\begin{equation}
  \eta_g=
  \begin{cases}
    \sin^2(\theta/2), & g=\mathrm{CP}(\theta),\\
    \sin^2(\theta), & g=R_{ZZ}(\theta),\\
    1, & g\in\{\mathrm{CX},\mathrm{CZ}\},\\
    0, & g=\mathrm{SWAP},
  \end{cases}
  \label{eq:gate-entanglement-capability}
\end{equation}
and
\begin{equation}
  \kappa_g=
  \begin{cases}
    1, & g\in\{\mathrm{CP},R_{ZZ},\mathrm{CX},\mathrm{CZ}\},\\
    3, & g=\mathrm{SWAP}.
  \end{cases}
  \label{eq:gate-implementation-cost}
\end{equation}
Single-qubit gates do not contribute to these boundary quantities. A common two-qubit entanglement-capability criterion gives the values in Eq.~\eqref{eq:gate-entanglement-capability}. Let $C_{\max}(U)$ denote the maximum concurrence that a two-qubit unitary $U$ can generate from a separable pure input~\cite{Kraus2001optimal,Goerz2015hybrid}. For a diagonal gate $U=\mathrm{diag}(e^{i\phi_{00}},e^{i\phi_{01}},e^{i\phi_{10}},e^{i\phi_{11}})$, the maximal reachable concurrence is $|\sin(\gamma_{\mathrm{nl}}/2)|$, where $\gamma_{\mathrm{nl}}=\phi_{00}-\phi_{01}-\phi_{10}+\phi_{11}$~\cite{Goerz2015hybrid}. We use the squared quantity $\eta_g=C_{\max}(U_g)^2$ as a normalized entanglement-capability score. Thus $\mathrm{CP}(\theta)$ has $\gamma_{\mathrm{nl}}=\theta$ and gives $\eta_g=\sin^2(\theta/2)$, whereas $R_{ZZ}(\theta)$ has $\gamma_{\mathrm{nl}}=-2\theta$ and gives $\eta_g=\sin^2(\theta)$. The fixed CZ gate is $\mathrm{CP}(\pi)$, and CX is locally equivalent to CZ, so both attain $\eta_g=1$. SWAP maps every product input $|a\rangle|b\rangle$ to the product state $|b\rangle|a\rangle$ and therefore has $\eta_g=0$. Squaring preserves the ordering and the zero/maximal endpoints of concurrence on $[0,1]$ while using the standard squared-concurrence scale~\cite{Coffman2000distributed}. Equation~\eqref{eq:gate-entanglement-capability} is therefore a gate-level entanglement-capability quantity; it is not the entanglement of the circuit state. Equation~\eqref{eq:gate-implementation-cost} separately counts native two-qubit-gate units under the fixed implementation model.

For a partition $P$, we define the normalized boundary contributions
\begin{align}
  E_{\mathrm{cut}}(P)
  &=\frac{\sum_{g\in\delta(P)}\eta_g}
          {\sum_{g\in\mathcal{G}_2(C)}\eta_g+\varepsilon},
  \label{eq:ep-static}\\
  I_{\mathrm{cut}}(P)
  &=\frac{\sum_{g\in\delta(P)}\kappa_g}
          {\sum_{g\in\mathcal{G}_2(C)}\kappa_g+\varepsilon},
  \label{eq:implementation-cost}
\end{align}
where $\varepsilon$ is used only to handle a zero-total denominator. A general gate-aware boundary score is
\begin{equation}
  \gatecutscore(P)
  =\lambda_E E_{\mathrm{cut}}(P)+\lambda_I I_{\mathrm{cut}}(P),
  \qquad \lambda_E+\lambda_I=1.
  \label{eq:gate-cut-score}
\end{equation}
In this work we use equal weights, $\lambda_E=\lambda_I=1/2$, and denote the resulting score by EP-Cost:
\begin{equation}
  \epcost(P)=\frac{1}{2}\left(E_{\mathrm{cut}}(P)+I_{\mathrm{cut}}(P)\right).
  \label{eq:ep-cost}
\end{equation}
EP-Cost combines the parameter-dependent entanglement-capability score with native implementation cost. For the parameterized CP and $R_{ZZ}$ gates used here, the same angle dependence also appears in the native single-gate QPD cost: the sampling-overhead factors are $(1+2|\sin(\theta/2)|)^2$ and $(1+2|\sin(\theta)|)^2$, respectively~\cite{Mitarai2021overhead}. Hence, for these gates the single-gate QPD overhead is a monotone function of the entanglement-capability score, $\Omega_g=(1+2\sqrt{\eta_g})^2$. This relation explains why the entanglement-capability term can distinguish equal-cut boundaries with different gate parameters, but it does not make EP-Cost a QPD objective. In particular, SWAP has zero entanglement capability under this product-input criterion although it is nontrivial to implement and cut; its implementation burden is represented separately through $\kappa_g$. The QPD coefficient norm $\gamma_g$ and aggregate $\Omega_{\mathrm{QPD}}$ remain reconstruction quantities and do not enter Eq.~\eqref{eq:ep-cost}.

\subsection{EP-Cost-ranked candidate screening}
SACE separates candidate generation from finite-bond refinement. Let $\mathcal{A}_{\mathrm{EP}}$ denote the archive of feasible partitions produced by a search procedure using EP-Cost as the boundary objective. Complement-equivalent bipartitions are canonicalized, duplicate partitions are removed, and the remaining archive is ranked deterministically by
\begin{equation}
  \bigl(\epcost(P),\,\operatorname{id}(P)\bigr),
  \label{eq:ep-rank-key}
\end{equation}
where $\operatorname{id}(P)$ is a canonical partition identifier used only to resolve exact score ties.

For a user-specified panel size $M$, the finite-bond stage operates on the $M$ highest-ranked unique candidates,
\begin{equation}
  \mathcal{P}_{M}
  =\operatorname{TopM}_{(C_{\mathrm{EP}},\operatorname{id})}
    \bigl(\mathcal{A}_{\mathrm{EP}}\bigr).
  \label{eq:topm-panel}
\end{equation}
The panel is fixed before MPS evaluation. Any search procedure may generate $\mathcal{A}_{\mathrm{EP}}$ under the same feasibility and ranking rules. In the primary evaluations, we fix $M=32$ as a pre-specified intermediate candidate-panel budget. Sensitivity to the panel size is assessed separately at 24 qubits using nested panels with $M\in\{8,16,32,64\}$.

\subsection{Finite-bond refinement and SACE selection}
For a candidate partition $P\in\mathcal{P}_{M}$, isolated local subcircuits are obtained by removing cross-partition two-qubit gates. For block $k$, let $\ket{\psi^{(k)}_{\chi}(P)}$ and $\ket{\psi^{(k)}_{2\chi}(P)}$ denote the normalized MPS states at bond dimensions $\chi$ and $2\chi$. We define the finite-bond overlap
\begin{equation}
  F^{(k)}_{\chi,2\chi}(P)
  =\left|\braket{\psi^{(k)}_{\chi}(P)}{\psi^{(k)}_{2\chi}(P)}\right|^2,
  \label{eq:multi-fidelity}
\end{equation}
and the partition-level convergence score
\begin{equation}
  \rmultichi(P)
  =\max_k\left[1-F^{(k)}_{\chi,2\chi}(P)\right].
  \label{eq:r-multi}
\end{equation}
The score is evaluated by normalized MPS tensor contractions and does not require statevector expansion. Smaller values indicate less change when the bond dimension is doubled. We use $\rmultichi$ as a finite-bond convergence proxy; it does not directly measure entanglement or general MPS complexity.

For target bond dimension $\chi$, SACE selects from the EP-Cost-ranked panel by
\begin{equation}
  P^{\star}_{\chi}
  =\underset{P\in\mathcal{P}_{M}}{\operatorname{arg\,min}}^{\mathrm{lex}}
  \left(\rmultichi(P),\,\epcost(P),\,\operatorname{id}(P)\right).
  \label{eq:sace-selector}
\end{equation}
The lexicographic rule prioritizes finite-bond convergence, then EP-Cost, with the canonical identifier resolving remaining ties. No additional weight is needed to combine boundary cost and finite-bond behavior.

\subsection{Reconstruction constraints, selection procedure, and computational cost}
Reconstruction resources are treated as external feasibility constraints rather than as terms in the SACE ranking objective. For a cut gate $g$, let $\gamma_g$ denote the one-norm of the selected QPD coefficient vector. The multiplicative QPD overhead is
\begin{equation}
  \qpdoverhead(P)=\prod_{g\in\delta(P)}\gamma_g^2,
  \label{eq:qpd-overhead}
\end{equation}
with the equivalent log-domain representation
\begin{equation}
  C_{\qpd}(P)=\log\qpdoverhead(P)
  =2\sum_{g\in\delta(P)}\log\gamma_g.
  \label{eq:qpd-cost}
\end{equation}

When an execution environment specifies a reconstruction budget
$B=(K_{\max},T_{\max},\Omega_{\max})$, the fixed candidate panel is restricted to
\begin{equation}
  \mathcal{P}_{M}^{\mathrm{feas}}(B)
  =\left\{P\in\mathcal{P}_{M}:\;
  K(P)\le K_{\max},\;
  T_{\mathrm{QPD}}(P)\le T_{\max},\;
  \qpdoverhead(P)\le\Omega_{\max}\right\},
  \label{eq:qpd-feasible-panel}
\end{equation}
where $K(P)$ is the number of cut gates and $T_{\mathrm{QPD}}(P)$ is the exact tensor-product term count implied by the selected QPD bases. SACE then applies the same lexicographic rule to the feasible subset,
\begin{equation}
  P_{\chi,\mathrm{QPD}}^{\star}
  =\underset{P\in\mathcal{P}_{M}^{\mathrm{feas}}(B)}{\operatorname{arg\,min}}^{\mathrm{lex}}
  \left(\rmultichi(P),\,\epcost(P),\,\operatorname{id}(P)\right).
  \label{eq:qpd-feasible-selector}
\end{equation}
The execution setting determines the budget; Eq.~\eqref{eq:qpd-feasible-selector} selects within the resulting feasible subset.

\paragraph{Selection procedure and information separation.}
Candidate generation and partition selection use only EP-Cost and finite-bond MPS overlaps. Exact-state quantities, reconstruction-quality metrics, and other validation diagnostics are evaluated after selection and do not affect the candidate panel or the selected partition.

\begin{figure}[!htb]
  \centering
  \includegraphics[width=\linewidth]{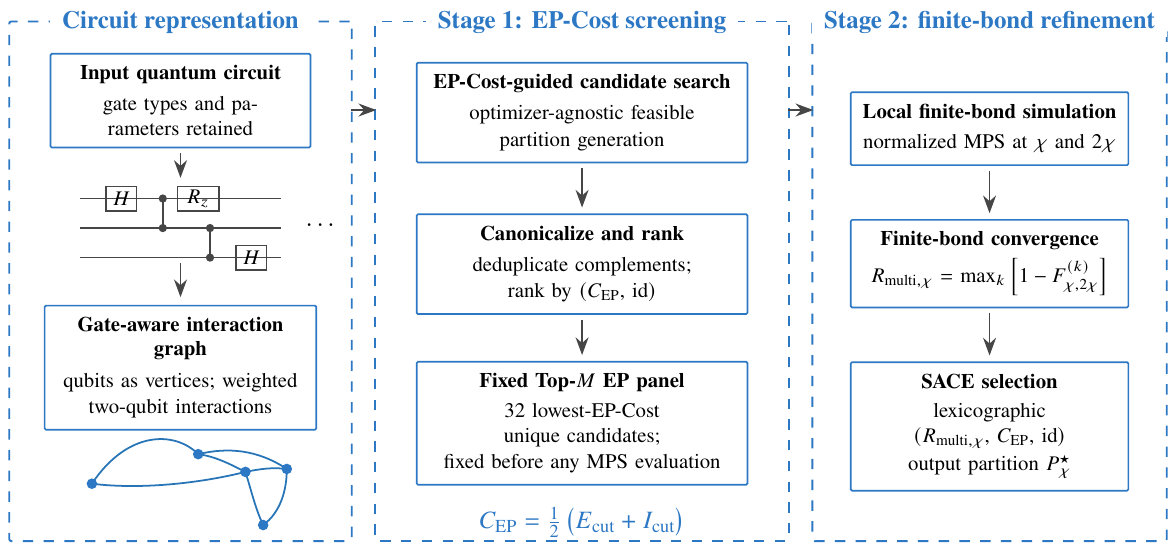}
  \caption{Overview of the SACE framework. Left: the input circuit is converted into a gate-aware interaction graph. Middle (Stage~1): EP-Cost guides candidate generation, canonicalization, and ranking to form a fixed Top-$M$ candidate panel ($M=32$ in the evaluation). Right (Stage~2): retained candidates are evaluated by finite-bond MPS simulations at $\chi$ and $2\chi$, and the final partition is selected lexicographically using $R_{\mathrm{multi},\chi}$, EP-Cost, and a canonical identifier. External reconstruction constraints may restrict the admissible candidate set but are omitted from the diagram for clarity.}
  \Description[Overview of the SACE framework]{A three-column workflow. The left column maps an input quantum circuit with retained gate types and parameters to a gate-aware interaction graph. The middle column shows Stage 1: EP-Cost-guided candidate search, canonicalization and ranking, and construction of a fixed Top-$M$ panel, with $M=32$ in the evaluation. The right column shows Stage 2: MPS simulation at bond dimensions $\chi$ and $2\chi$, computation of $R_{\mathrm{multi},\chi}$, and lexicographic selection of the final partition.}
  \label{fig:framework}
\end{figure}

Figure~\ref{fig:framework} shows the two-stage selection workflow. Optional reconstruction constraints are specified in Eq.~\eqref{eq:qpd-feasible-panel} and Algorithm~\ref{alg:selection}.

Algorithm~\ref{alg:selection} summarizes the candidate-level SACE procedure.

\begin{algorithm}
\caption{Two-stage SACE partition selection}
\label{alg:selection}
\begin{algorithmic}[1]
\REQUIRE Circuit $C$, block-capacity constraints, target bond dimension $\chi$, panel size $M$, feasible candidate-generation procedure, optional reconstruction budget $B$
\ENSURE Selected partition $P^{\star}_{\chi}$ (or $P^{\star}_{\chi,\mathrm{QPD}}$ when $B$ is applied), or terminate if no valid candidate set is available
\STATE Build the gate-aware interaction graph $G_C$
\STATE Generate a feasible candidate archive $\mathcal{A}_{\mathrm{EP}}$ using EP-Cost
\STATE Complement-canonicalize and deduplicate the archive
\STATE Sort unique candidates by $(\epcost,\operatorname{id})$
\IF{fewer than $M$ unique feasible candidates are available}
  \STATE \textbf{terminate without a valid selection}
\ENDIF
\STATE Fix $\mathcal{P}_{M}$ as the first $M$ ranked candidates
\IF{an external reconstruction budget $B$ is specified}
  \STATE Restrict $\mathcal{P}_{M}$ to $\mathcal{P}_{M}^{\mathrm{feas}}(B)$
  \IF{$\mathcal{P}_{M}^{\mathrm{feas}}(B)=\varnothing$}
    \STATE \textbf{terminate without a valid selection}
  \ENDIF
\ENDIF
\FOR{each retained candidate $P$}
  \STATE Extract the isolated local subcircuits
  \STATE Simulate normalized MPS states at bond dimensions $\chi$ and $2\chi$
  \STATE Compute $\rmultichi(P)=\max_k[1-F^{(k)}_{\chi,2\chi}(P)]$
\ENDFOR
\STATE Select lexicographically by $(\rmultichi,\epcost,\operatorname{id})$
\STATE \textbf{return} the selected partition
\end{algorithmic}
\end{algorithm}

\paragraph{Computational cost.}
Candidate generation evaluates EP-Cost without MPS simulation. Finite-bond refinement simulates each local block at $\chi$ and $2\chi$ for candidates in the fixed Top-$M$ panel, or its budget-feasible subset. Identical local-subcircuit results can be cached and reused. Exact-state validation and QPD reconstruction are performed after selection and are excluded from the selection cost.

\section{Evaluation Methodology}

Across all evaluations, candidate generation and SACE selection are completed before post-selection accuracy or reconstruction metrics are computed. Unless otherwise stated, the finite-bond stage uses $M=32$ in Eq.~\eqref{eq:topm-panel}.

\subsection{Benchmarks, baselines, and common evaluation setup}
QFT tests parameter-dependent CP cut costs, while weighted-QAOA tests native $R_{ZZ}$ interactions on sparse graphs. The small-scale candidate-space study includes full QFT-8 and QFT-9, approximate-QFT variants at truncation depths $d=3$ and $d=2$, and eight-qubit weighted-QAOA circuits on Star, Erd\H{o}s--R\'enyi, Small-world, Cycle, 3-regular, Path, $2\times4$ grid, and Two-community graphs~\cite{NielsenChuang2010,Farhi2014original,Farhi2014qaoa}.

Complement-equivalent bipartitions are canonicalized before evaluation. The eight-qubit spaces contain 91 feasible 3/5 or 4/4 partitions, and the nine-qubit spaces contain 210 feasible 3/6 or 4/5 partitions, allowing exhaustive evaluation of EP-Cost and native QPD overhead. Six representative circuits are used for sampled-QPD reconstruction: QFT-8, QFT-9, Star, Erd\H{o}s--R\'enyi, Cycle, and 3-regular. The remaining circuits are used for candidate-space and feasibility analyses.

For each small-scale circuit, the EP-Cost-guided genetic algorithm (GA) is compared with three graph-based controls: an exact unit-cut oracle, Kernighan--Lin (KL)~\cite{Kernighan1970}, and spectral partitioning~\cite{Ng2002spectral}. The unit-cut oracle is obtained by exhaustive enumeration of the feasible candidate space and minimizes the number of cross-partition two-qubit gates; ties are resolved by smaller maximum block size and then by the canonical partition identifier. Exhaustive enumeration also identifies the EP-Cost co-optimal set and the native-QPD-overhead rank of each selected partition. The GA uses seeds 0, 1, and 2, giving three independent search runs per candidate space. EP-Cost co-optimality and native-QPD-overhead ranks use an absolute tolerance of $10^{-12}$: a candidate is co-optimal when its score differs from the enumerated minimum by at most this tolerance, and the overhead rank is one plus the number of candidates whose overhead is smaller by more than this tolerance.

The exhaustive spaces also define a native-QPD-overhead oracle as the co-optimal set that minimizes $\Omega_{\mathrm{QPD}}$ over all legal partitions. Oracle membership is evaluated with relative and absolute tolerances of $10^{-12}$, so numerically tied partitions are retained rather than reduced to a single identifier. We compare both the GA-selected partition and the exact EP-Cost optimum with this oracle. This analysis tests empirical alignment between EP-Cost and native QPD overhead; it does not treat $\Omega_{\mathrm{QPD}}$ as part of the EP-Cost objective.

The implementation uses Qiskit 2.4.1, Qiskit Aer 0.17.2, and Qiskit Addon Cutting 0.10.0~\cite{JavadiAbhari2024qiskit,QiskitAddonCutting2025}. For reproducibility, we retain machine-readable circuit definitions, candidate sets, method selections, validation outcomes, sampled-QPD tasks, aggregate metrics, and explicit failure or resource-limit states. These records are covered by the Data and Code Availability statement.

\subsection{Small-scale sampled-QPD evaluation}
Sampled-QPD reconstruction uses the same resource limits across both circuit types unless otherwise stated: 1,000 requested QPD samples, 1,024 shots per generated subexperiment, at most 20 gate cuts, a native-QPD-overhead cap of $10^6$, and at most 20,000 generated subexperiments. Subexperiments are simulated with Qiskit Aer in MPS mode at maximum bond dimension 64 and zero truncation threshold. The exact input-state and sampling seeds are retained with the reproducibility records.

QFT-8 and QFT-9 each use three fixed input states crossed with three fixed sampling seeds. The primary controls are spectral partitioning for QFT-8 and KL for QFT-9. Paired methods use equal cut counts to compare parameter-aware boundary selection and its effects on QPD overhead and reconstruction quality.

For weighted-QAOA, Star, Erd\H{o}s--R\'enyi, Cycle, and 3-regular use three paired sampling seeds, with the unit-cut oracle as the primary control. In addition to distribution-level total-variation distance and Hellinger metrics, QAOA reconstruction is evaluated through weighted Max-Cut expectation error. These cases provide complementary evidence because application-specific observable error need not vary monotonically with QPD overhead.

For reference distribution $p$ and reconstructed physical distribution $\widehat p$, total-variation distance (TVD) and Hellinger fidelity are
\begin{equation}
  D_{\mathrm{TV}}(p,\widehat p)
  =\frac{1}{2}\sum_x\left|p_x-\widehat p_x\right|,
  \label{eq:tvd}
\end{equation}
\begin{equation}
  \begin{aligned}
    \hellingerfid(p,\widehat p)
    &=\left(\sum_x\sqrt{p_x\widehat p_x}\right)^2,\\
    \hellingerinf(p,\widehat p)
    &=1-\hellingerfid(p,\widehat p).
  \end{aligned}
  \label{eq:hellinger}
\end{equation}
Lower TVD and lower Hellinger infidelity indicate better reconstruction. Negative quasiprobability mass before projection is reported only as a reconstruction diagnostic.

For the QAOA circuits, reconstruction is also evaluated through the weighted Max-Cut expectation
\begin{equation}
  \expect{C}
  =\sum_{(u,v)\in E}\frac{w_{uv}}{2}
   \left(1-\expect{Z_uZ_v}\right),
  \qquad
  \epsilon_C=\left|\expect{C}_{\mathrm{rec}}-\expect{C}_{\mathrm{ref}}\right|.
  \label{eq:qaoa-cost-error}
\end{equation}
The circuit instance, rather than an individual sampling repeat, is the statistical unit. Repeat-level means and sample standard deviations characterize finite-sampling variability within each circuit.

A complementary 24-qubit sparse weighted-QAOA case examines the same cut mechanism at a larger width. It compares two strict 12/12 partitions with four native $R_{ZZ}$ cuts and equal implementation-cost contribution. Because this case reconstructs edge marginals rather than the full 24-qubit joint distribution, it is analyzed separately from the small-scale sampled-QPD benchmark.

\subsection{Reconstruction and finite-bond validation}
We evaluate complete deterministic reconstruction at 12 qubits and local finite-bond accuracy or consistency at larger widths.
\paragraph{Deterministic-QPD reconstruction at 12 qubits.}
\label{sec:aqft-qpd-bridge-method}
The AQFT-12 study evaluates complete deterministic-QPD reconstruction under explicit feasibility limits. An additional five-cut comparison controls cut count and QPD term count. The circuit uses truncation distance $d=2$, retaining 21 CP gates between qubits separated by at most two positions. No physical SWAP gates are inserted; bit reversal affects only output ordering. Three fixed product-state inputs use per-qubit $R_Y(\theta)$ and $R_Z(\phi)$ rotations on the same topology, with generating seeds retained in the reproducibility records.

All partitions are strict $6|6$ bipartitions. Target bond dimensions are $\chi\in\{2,4,8\}$, with $\chi=8$ serving as an exact local MPS control for six-qubit blocks. SACE uses the same Top-32 EP-Cost-ranked candidate protocol as Eq.~\eqref{eq:topm-panel}. Unweighted KL and spectral clustering select the same physical partition in every comparison and are therefore represented as a shared KL/Spectral control. This comparison includes no EP-only baseline.

\begin{table}[!htb]
  \caption{Feasibility constraints for the 12-qubit AQFT ($d=2$) deterministic-QPD validation.}
  \label{tab:aqft-qpd-bridge-constraints}
  \small
  \setlength{\tabcolsep}{4pt}
  \begin{tabular}{>{\raggedright\arraybackslash}p{0.25\linewidth}>{\raggedright\arraybackslash}p{0.20\linewidth}>{\raggedright\arraybackslash}p{0.45\linewidth}}
    \toprule
    Constraint & Value & Rationale \\
    \midrule
    Circuit & 12-qubit AQFT ($d=2$) & Tractable full deterministic-QPD bridge with finite-bond variation \\
    Local blocks & strict $6|6$ & Fixed local simulation width \\
    Input states & three fixed product states & One fixed AQFT topology with reproducible input generation \\
    Candidate panel & Top-32 EP-Cost-ranked & Same candidate protocol as SACE \\
    Target bond dimensions & $\{2,4,8\}$ & Constrained, intermediate, and exact six-qubit local controls \\
    Maximum cut gates & $K_{\max}=5$ & Per-task deterministic-enumeration budget \\
    Maximum exact terms & $T_{\max}=7776=6^5$ & Six QPD basis terms per cut CP gate \\
    Maximum overhead & $\Omega_{\max}=2000$ & Non-binding safeguard within the retained five-cut tier \\
    Implied coefficient norm & $\Gamma\le\sqrt{2000}\approx44.72$ & Because $\Omega=\Gamma^2$ \\
    Reconstruction & complete deterministic enumeration & No QPD sampling or finite-shot noise \\
    \bottomrule
  \end{tabular}
\end{table}

Exhaustive enumeration of the 462 complement-unique strict $6|6$ partitions leaves four feasible partitions under Table~\ref{tab:aqft-qpd-bridge-constraints}, all within Top-32. With six QPD terms per CP cut, the cut-count and term-count constraints are equivalent. The smallest tractable tier admitting multiple partitions is $K_{\max}=5$; the overhead cap is non-binding. These budgets specify the evaluation setting, not universal SACE constants.

We evaluate all three feasible five-cut partitions over the six conditions formed by three input states and $\chi\in\{2,4\}$. Each candidate uses 7,776 deterministic-QPD terms; $\Gamma$ and $\Omega_{\mathrm{QPD}}$ remain partition dependent. Reconstruction and error calculations are unchanged.

In this deterministic setting, each retained cut CP gate has six QPD basis terms, so the complete tensor-product enumeration size is
\begin{equation}
  T_{\mathrm{QPD}}(P)
  =\prod_{g\in\delta(P)}m_g
  =6^{K(P)}.
  \label{eq:aqft-qpd-terms}
\end{equation}
The relation $T_{\mathrm{QPD}}=6^K$ applies to the CP-gate bases used here. Reconstruction enumerates every tensor term without QPD sampling or finite-shot noise. The exact full-circuit statevector is used only for post-selection evaluation.

TVD and Hellinger infidelity use Eqs.~\eqref{eq:tvd} and~\eqref{eq:hellinger} on the simplex-projected physical distribution. To retain information before projection, we additionally define the raw signed-distribution $L_1$ error
\begin{equation}
  E_{\mathrm{raw}\text{-}L1}
  =\left\|p_{\mathrm{exact}}-\widetilde p_{\mathrm{signed}}\right\|_1,
  \label{eq:aqft-raw-l1}
\end{equation}
where $\widetilde p_{\mathrm{signed}}$ is the signed deterministic-QPD reconstruction before projection.

The observable set contains 101 Pauli operators: 36 single-qubit $X/Y/Z$ observables, 63 $XX/YY/ZZ$ observables associated with the 21 retained AQFT CP edges, and two global $X$- and $Z$-parity observables. Their mean absolute error (MAE) is
\begin{equation}
  E_{\mathrm{Pauli}}
  =\frac{1}{|\mathcal O|}\sum_{O\in\mathcal O}
  \left|\langle O\rangle_{\mathrm{rec}}-\langle O\rangle_{\mathrm{exact}}\right|,
  \qquad |\mathcal O|=101.
  \label{eq:aqft-pauli-mae}
\end{equation}

Finally, we use the coefficient-normalized local finite-bond diagnostic
\begin{equation}
  E_{\mathrm{loc},\chi}^{\mathrm{norm}}
  =\frac{1}{\Gamma}\sum_j |c_j|\,\delta_{j,\chi},
  \qquad
  \delta_{j,\chi}
  =\max_{b\in\{A,B\}}
  \left\|\rho_{j,b}^{(\chi)}-\rho_{j,b}^{(8)}\right\|_{F},
  \label{eq:aqft-local-mps-error}
\end{equation}
where $j$ indexes QPD tensor terms, $c_j$ is the QPD coefficient, $b$ indexes the two local blocks, and $\rho_{j,b}^{(\chi)}$ is the finite-bond local signed density-matrix contribution. The $\chi=8$ quantity is the six-qubit local control. This post-selection diagnostic measures coefficient-normalized local approximation error in the QPD terms; it differs from the isolated-subcircuit selection score and is not a bound on global reconstruction error.

\paragraph{Exact-local finite-bond validation at 24 qubits.}
For each Top-32 candidate, SACE computes $\rmultichi$ from isolated local MPS simulations at $\chi$ and $2\chi$ using Eq.~\eqref{eq:r-multi}. Post-selection validation compares the target-$\chi$ states with exact isolated-subcircuit states. For block $k$,
\begin{equation}
  F^{(k)}_{\mathrm{exact},\chi}(P)
  =\left|\braket{\psi^{(k)}_{\mathrm{exact}}(P)}{\psi^{(k)}_{\chi}(P)}\right|^2,
  \label{eq:exact-local-fidelity}
\end{equation}
and the validation loss is the worst-block infidelity
\begin{equation}
  \localinf(P)
  =\max_k\left[1-F^{(k)}_{\mathrm{exact},\chi}(P)\right].
  \label{eq:local-validation-loss}
\end{equation}
Within the fixed panel, the post-selection oracle is
\begin{equation}
  P^{\mathrm{oracle}}_{\chi}
  =\arg\min_{P\in\mathcal{P}_{32}}\localinf(P),
\end{equation}
and the selector-to-oracle gap is
\begin{equation}
  \Delta^{\mathrm{oracle}}_{\chi}
  =\localinf(P^{\star}_{\chi})-\localinf(P^{\mathrm{oracle}}_{\chi})\geq0.
  \label{eq:selector-oracle-gap}
\end{equation}
Spearman association between $\rmultichi$ and $\localinf$ over the fixed panel is used as a ranking diagnostic. These quantities characterize isolated-subcircuit MPS accuracy and are not interpreted as global-circuit or QPD-reconstruction fidelity.

Held-out validation uses six previously unseen 24-qubit circuits: three QFT instances and three weighted-QAOA instances generated from random seeds disjoint from the development sets. The exact seed values are retained with the reproducibility records rather than used as manuscript-level instance names. Each QFT instance preserves the same two-qubit interaction pattern while varying the initialization parameters. Each weighted-QAOA instance contains 48 unique weighted problem edges with fixed $\gamma=\beta=1$. All partitions are strict 12/12 splits.

For each held-out instance, candidate generation uses three EP-Cost-guided GA searches with 2,000 objective evaluations per run. The complement-deduplicated archive is ranked by Eq.~\eqref{eq:ep-rank-key}, and the Top-32 panel is fixed before MPS evaluation. Target bond dimensions are $\chi\in\{2,4,8\}$ with references $2\chi\in\{4,8,16\}$, producing 18 instance--$\chi$ comparisons. SACE is compared with EP-only, KL, and spectral partitioning.

We also report the selector--oracle gap and the rank association between $\rmultichi$ and exact-local error, separately for QFT and weighted-QAOA.

\subsection{Large-scale and optimizer-robustness evaluation}
The initial 24-qubit validation uses six circuits, three QFT and three weighted-QAOA, while a separate benchmark uses twenty instances, ten of each type. Both use exact isolated-subcircuit references. At 50 qubits, ten QFT and nine weighted-QAOA instances use higher-resource MPS references with explicit convergence diagnostics. QFT provides the primary scale-up evidence; weighted-QAOA is a reference-limited stress test. Benchmark seeds are disjoint from the development and initial validation sets.

The optimizer study reuses five QFT and five weighted-QAOA instances from the 24-qubit benchmark. It retains their frozen GA results and evaluates simulated annealing (SA) and discrete particle swarm optimization (DPSO) on the same circuits.

The large-scale benchmark retains the Top-32 panel, EP-Cost ranking, $\rmultichi$ definition, and lexicographic tie breakers. Partitions are strict 12/12 splits at 24 qubits and 25/25 splits at 50 qubits, with $\chi\in\{2,4,8\}$. At each width, QFT instances share the two-qubit gate structure and vary only the per-qubit $R_Y$ and $R_Z$ product-state initialization parameters.

Panel-size sensitivity is evaluated on the twenty 24-qubit benchmark instances using the same frozen, complement-deduplicated GA archives. The ordered archives define strictly nested Top-$M$ panels for $M\in\{8,16,32,64\}$. Existing Top-32 candidate results are reused, and ranks 33--64 are evaluated with the same finite-bond and exact-local procedures. For each instance--$\chi$ cell, the analysis also records whether Top-$M$ contains a Top-64 co-optimum under $\rmultichi$ and, independently, under $Q_{\mathrm{exact},\chi}$. The latter is an oracle only within the evaluated Top-64 panel and is not a global partition-space optimum. Co-optimality in this sensitivity analysis uses relative tolerance $10^{-9}$ and absolute tolerance $10^{-12}$.

At 24 qubits, the validation metric is the exact-local worst-block infidelity already defined in Eq.~\eqref{eq:local-validation-loss}, denoted here by $Q_{\mathrm{exact},\chi}(P)=\localinf(P)$. At 50 qubits, validation compares each normalized local MPS state at $\chi$ with a reference at $4\chi$:
\begin{equation}
 Q_{\mathrm{highref},\chi}(P)
 =\max_k\left[1-\left|\braket{\psi^{(k)}_{\chi}(P)}{\psi^{(k)}_{4\chi}(P)}\right|^2\right].
 \label{eq:highref-validation}
\end{equation}
Reference convergence is assessed by
\begin{equation}
 A_{\mathrm{ref},\chi}(P)
 =\max_k\left[1-\left|\braket{\psi^{(k)}_{2\chi}(P)}{\psi^{(k)}_{4\chi}(P)}\right|^2\right].
 \label{eq:reference-adequacy}
\end{equation}
All three quantities use the worst local block. The selector $\rmultichi$ compares $\chi$ with $2\chi$; Eqs.~\eqref{eq:highref-validation} and~\eqref{eq:reference-adequacy} are evaluated after selection and use the additional $4\chi$ reference. Smaller values indicate closer agreement. The 50-qubit metric measures consistency with a finite-resource reference, so its interpretation depends on $A_{\mathrm{ref},\chi}$; it does not establish exact-state accuracy. Tables report $A_{\mathrm{ref},\chi}$ for the SACE-selected partition. Runtime is separated into candidate generation, finite-bond probing, and post-selection validation.

On the ten-instance 24-qubit subset, GA, SA~\cite{Kirkpatrick1983sa}, and DPSO~\cite{Kennedy1995pso} each run three searches with seeds 42, 43, and 44 and 2,000 EP-objective evaluations per seed, totaling 6,000 per optimizer and instance. Visited partitions are pooled and complement-deduplicated before fixing Top-32. GA reuses its benchmark searches; SA and DPSO generate their own archives under the same 12/12 constraints. All use exact-local validation at $\chi\in\{2,4,8\}$. Pairwise Jaccard overlap and selected-partition equality describe search diversity and do not affect selection.

For the large-scale and optimizer studies, local qubits retain ascending original-qubit order. Win/loss/tie comparisons use an absolute error tolerance of $10^{-12}$. The large-scale benchmark reports 95\% percentile bootstrap intervals for median paired improvements, using 10,000 resamples of circuit instances and retaining their associated bond-dimension conditions together.

\section{Results}
\label{sec:results}

We first examine the screening stage over the 14 exhaustively enumerated small-scale candidate spaces. The EP-Cost-guided GA selection belongs to the global $\Omega_{\mathrm{QPD}}$ co-optimal set in all 14 spaces, and the exact EP-Cost optimum intersects the same set in all 14. The GA-to-oracle overhead ratio is therefore 1 throughout. Across the complete candidate landscapes, the median Spearman and Kendall $\tau_b$ associations between EP-Cost and native QPD overhead are 0.9965 and 0.9619, respectively, with ranges of 0.940--1.000 and 0.805--0.996. These results show close agreement between EP-Cost rankings and native QPD overhead across the tested candidate spaces. Per-circuit correlations and tie-aware QPD-oracle ranks are reported in Appendix Fig.~\ref{fig:small-scale-candidate-space}.

\begin{figure}[!htbp]
  \centering
  \includegraphics[width=\linewidth,keepaspectratio]{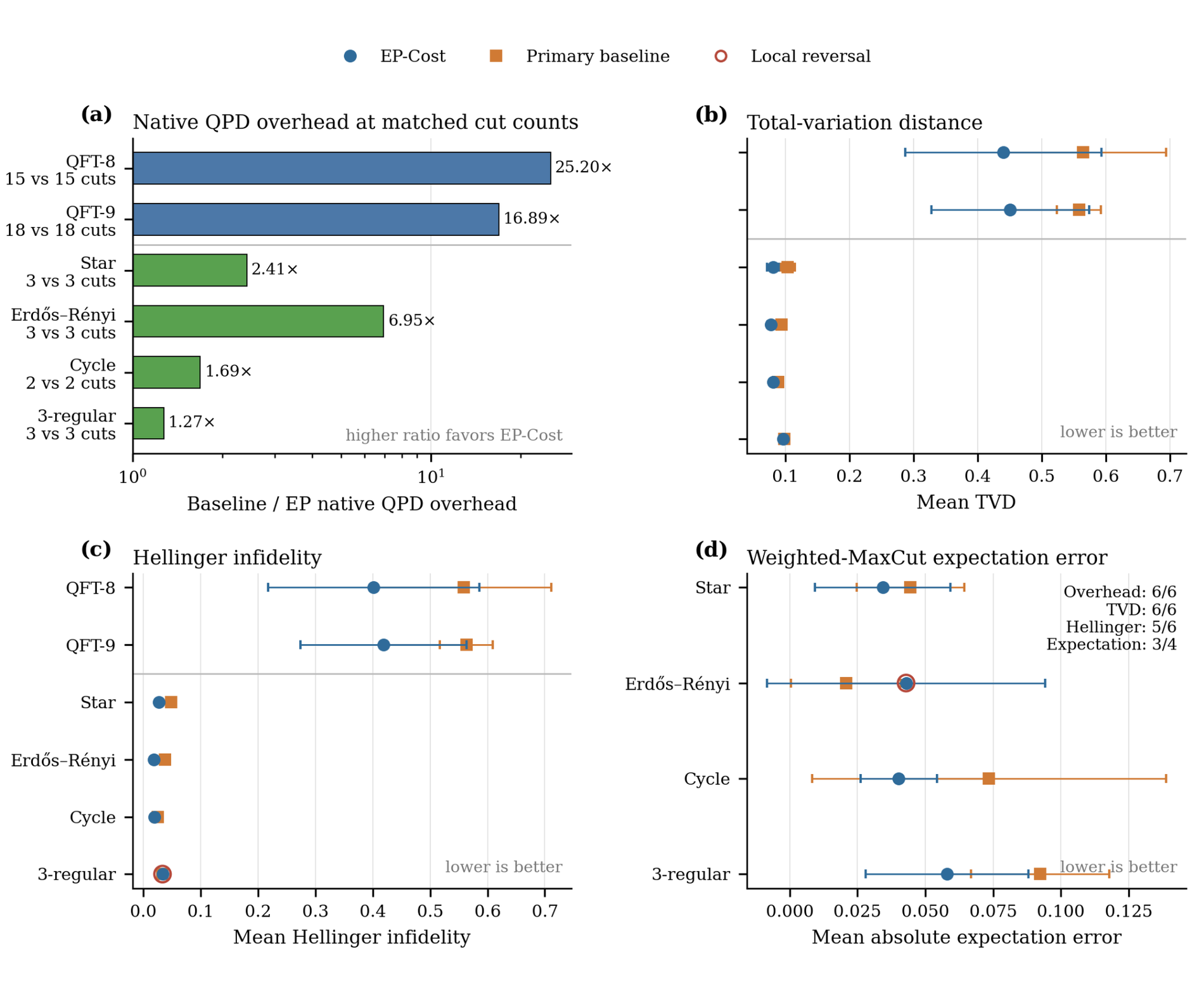}
  \caption{Small-scale sampled-QPD reconstruction across six benchmark circuits. (a) Baseline-to-EP-Cost native QPD-overhead ratios; values greater than one favor EP-Cost. Every comparison has matched cut count. (b) Mean total-variation distance and (c) mean Hellinger infidelity for EP-Cost and the primary baseline; error bars show within-circuit sample standard deviations. (d) Mean weighted-Max-Cut absolute expectation error for the four QAOA circuits. Hollow red circles mark local metric reversals. The analysis unit is the circuit instance, not the repeat seed.}
  \Description[Small-scale sampled-QPD reconstruction]{A four-panel figure compares EP-Cost-guided GA with primary baselines on QFT-8, QFT-9, Star, Erd\H{o}s--R\'enyi, Cycle, and 3-regular circuits. The panels show native QPD-overhead ratios, total-variation distance, Hellinger infidelity, and weighted-Max-Cut expectation error. The overhead ratios favor EP-Cost across all six circuits, while the reconstruction panels contain two highlighted local reversals.}
  \label{fig:small-scale-end-to-end}
\end{figure}

The sampled-QPD study then tests whether this screening behavior is reflected in reconstructed quantities. Most graph-based controls recover the same canonical partition as EP-Cost or an analytically equivalent solution; six distinct, feasible benchmark comparisons are retained in Table~\ref{tab:small-scale-confirmatory-results} and Fig.~\ref{fig:small-scale-end-to-end}. QFT provides the main validation chain, while weighted-QAOA is used as a complementary stress test.

\begin{table}[!htbp]
\centering
\footnotesize
\caption{Small-scale sampled-QPD results for six benchmark circuits. Analytical partition quantities and sampled reconstruction metrics are reported together for the same EP-Cost--baseline pairs. EP/Base denotes EP-Cost-guided GA and the primary baseline. All comparisons have equal cut counts. Lower TVD, Hellinger infidelity $I_H$, and Max-Cut error are better; a larger $\Omega_{\mathrm{base}}/\Omega_{\mathrm{EP}}$ favors EP-Cost. Full mean $\pm$ sample-standard-deviation values are given in Table~\ref{tab:small-scale-full-sampled-metrics}.}
\label{tab:small-scale-confirmatory-results}
\begingroup
\setlength{\tabcolsep}{4.2pt}
\renewcommand{\arraystretch}{1.04}
\begin{tabular*}{\linewidth}{@{\extracolsep{\fill}}llccccccc@{}}
\toprule
& & \multicolumn{4}{c}{Partition and QPD cost} & \multicolumn{3}{c}{Reconstruction quality} \\
\cmidrule(lr){3-6}\cmidrule(lr){7-9}
Circuit & Baseline & Split & $E_{\mathrm{cut}}$ & EP-Cost & $\Omega_{\mathrm{base}}/\Omega_{\mathrm{EP}}$ & TVD & $I_H$ & Max-Cut $|\Delta C|$ \\
 &  & EP/Base & EP/Base & EP/Base &  & EP/Base & EP/Base & EP/Base \\
\midrule
QFT-8 & Spectral & 3/5--3/5 & .205/.352 & .370/.444 & 25.20$\times$ & .440/.565 & .402/.558 & -- \\
QFT-9 & KL & 3/6--3/6 & .178/.300 & .339/.400 & 16.89$\times$ & .451/.558 & .419/.563 & -- \\
Star & Unit-cut & 3/5--3/5 & .256/.446 & .342/.437 & 2.41$\times$ & .080/.103 & .028/.049 & .034/.045 \\
Erd\H{o}s--R\'enyi & Unit-cut & 4/4--4/4 & .132/.425 & .216/.363 & 6.95$\times$ & .077/.094 & .019/.038 & .043/.021 \\
Cycle & Unit-cut & 3/5--4/4 & .139/.244 & .195/.247 & 1.69$\times$ & .081/.089 & .020/.025 & .040/.074 \\
3-regular & Unit-cut & 3/5--3/5 & .214/.239 & .232/.244 & 1.27$\times$ & .097/.098 & .034/.034 & .058/.092 \\
\bottomrule
\end{tabular*}
\endgroup
\end{table}

\subsection{QFT validation across reconstruction and scale}
At matched cut counts, EP-Cost reduces native QPD overhead by $25.20\times$ for QFT-8 and $16.89\times$ for QFT-9. Mean TVD falls from 0.565 to 0.440 and from 0.558 to 0.451, respectively; Hellinger infidelity falls from 0.558 to 0.402 and from 0.563 to 0.419. Since cut counts are identical, the comparison isolates differences in the selected parameterized CP boundaries.

\begin{figure}[!htb]
  \centering
  \includegraphics[width=\linewidth,keepaspectratio]{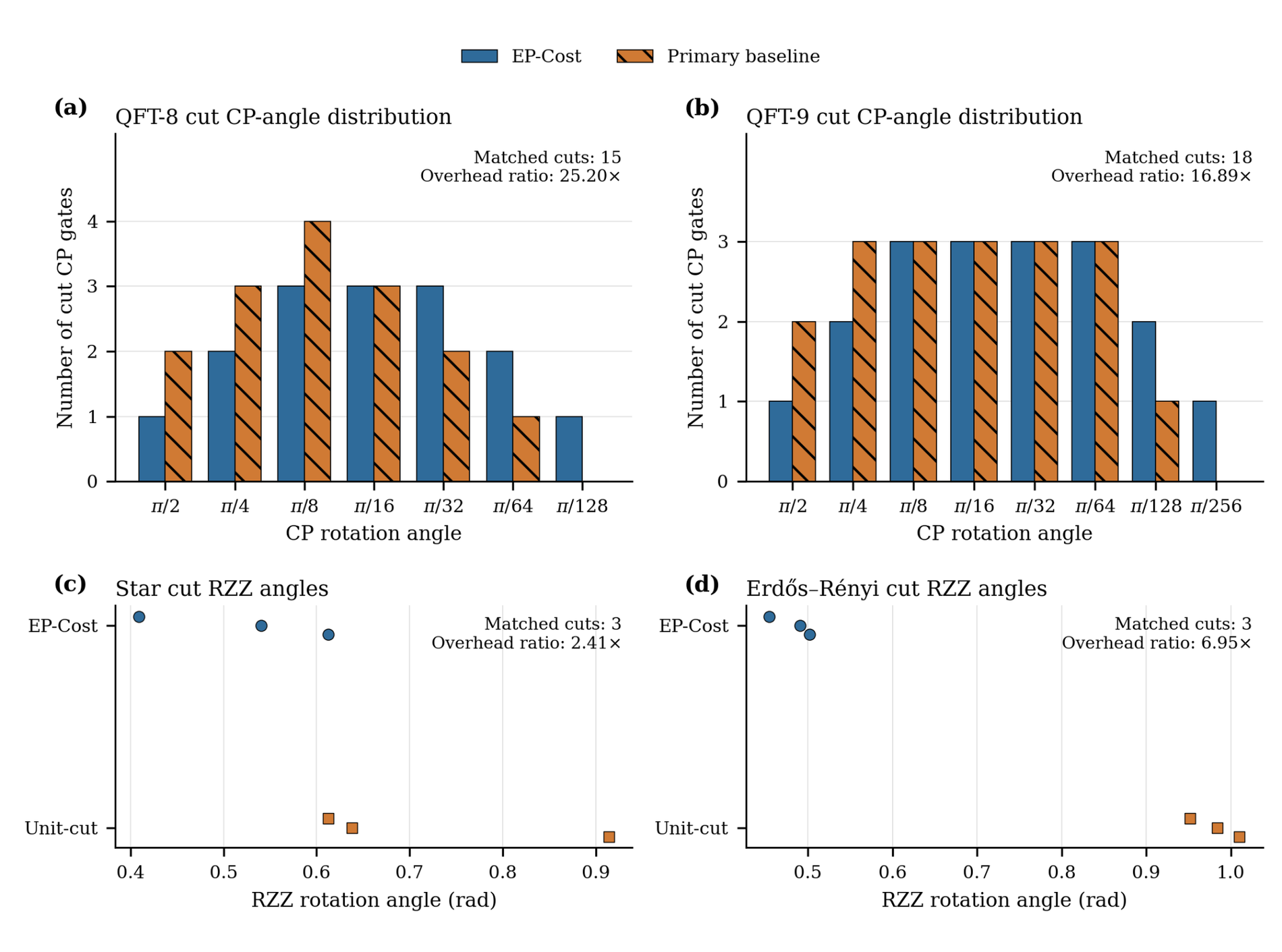}
  \caption{Entanglement-capability-aware matched-cut comparison. Panels (a) and (b) show the CP angles of the cut gates for QFT-8 and QFT-9, and panels (c) and (d) show the corresponding native $R_{ZZ}$ angles for Star and Erd\H{o}s--R\'enyi. EP-Cost and the primary baseline use the same cut count and implementation-cost contribution in each comparison; the reported ratios give the corresponding native QPD-overhead reduction.}
  \Description[Entanglement-capability-aware matched-cut mechanism]{Four panels compare the parameters of cross-partition gates selected by EP-Cost and primary baselines. The two QFT panels show histograms of controlled-phase angles, and the two QAOA panels show strip plots of RZZ angles. Each comparison has matched cut count and reports the corresponding baseline-to-EP-Cost QPD-overhead ratio.}
  \label{fig:small-scale-parameter-mechanism}
\end{figure}

\FloatBarrier
Figure~\ref{fig:small-scale-parameter-mechanism}(a,b) shows that EP-Cost selects boundaries with fewer large controlled-phase rotations than the graph-based controls. The resulting differences in gate-level entanglement capability and QPD cost explain why equal cut counts can incur different reconstruction overheads.

The 12-qubit AQFT study connects the finite-bond selector to complete deterministic reconstruction. In all six low-bond conditions, the SACE-selected partitions have lower coefficient-normalized local QPD-term error, with mean reductions of 33.1\% at $\chi=2$ and 49.2\% at $\chi=4$. Reconstruction errors also decrease: TVD, Hellinger infidelity, and raw $L_1$ are lower in five of six conditions, while Pauli-observable MAE is lower in all six. Because the QPD expansion is enumerated deterministically, these comparisons are not affected by sampling variability.

\begin{figure}[!htb]
  \centering
  \includegraphics[width=\linewidth,keepaspectratio]{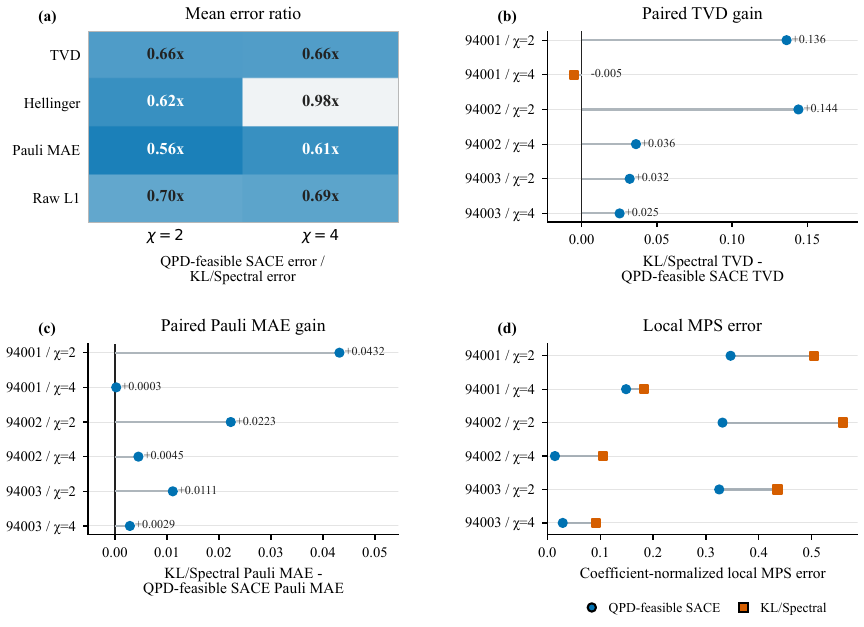}
  \caption{End-to-end deterministic-QPD reconstruction on 12-qubit AQFT ($d=2$). (a) Mean SACE-to-KL/Spectral error ratios at $\chi=2$ and $4$ for TVD, Hellinger infidelity, Pauli-observable MAE, and raw $L_1$ error. (b) Paired TVD improvements and (c) paired Pauli-MAE improvements, defined as control error minus SACE error, so positive values favor SACE. (d) Coefficient-normalized local finite-bond error from Eq.~\eqref{eq:aqft-local-mps-error}. KL and spectral clustering select the same physical control partition.}
  \Description[Deterministic-QPD bridge results on a 12-qubit AQFT with truncation distance two]{Four panels compare QPD-feasible SACE with a shared KL/spectral control at bond dimensions 2 and 4. The panels show mean error ratios, paired TVD gains, paired Pauli-observable mean-absolute-error gains, and coefficient-normalized local MPS errors across three product input states.}
  \label{fig:aqft-qpd-bridge-main}
\end{figure}

One condition at $\chi=4$ has a control-minus-SACE TVD difference of $-0.004933$; at $\chi=8$, all three methods select the same partition. At low bond dimensions, SACE uses five CP cuts and 7,776 QPD terms, compared with three cuts and 216 terms for KL/Spectral, giving an accuracy--resource trade-off. Among the three five-cut partitions, SACE is best on all four reconstruction metrics in four of six conditions, with median rank 1 and mean rank 1.5 for each metric. It is not optimal in the remaining two conditions. Appendix Fig.~\ref{fig:aqft-matched-resource-accuracy} and Table~\ref{tab:aqft-matched-resource-summary} report the matched-resource and Pareto results.

The QFT evidence remains consistent as the validation scale increases. At 24 qubits, where exact isolated-subcircuit states are available, SACE improves all ten QFT instances against EP-only, KL, and spectral partitioning at each of $\chi=2,4,8$, giving 30/30 favorable comparisons against every baseline. Median exact-local infidelity decreases from 0.3292 to 0.0597 and 0.0024 as $\chi$ increases. At 50 qubits, the validation reference is instead a finite-resource $4\chi$ MPS state. For QFT, its adequacy improves markedly: median $A_{\mathrm{ref},\chi}$ falls from 0.4274 to 0.1030 and 0.0167, while the median rank association between $\rmultichi$ and higher-resource validation rises from 0.234 to 0.822 and 0.940. The 50-qubit QFT results are therefore used as the primary scale-up consistency evidence rather than as reconstructed global-circuit fidelity.

\FloatBarrier
\subsection{Weighted-QAOA as a complementary stress test}
\label{sec:qaoa-results}
Weighted-QAOA provides a complementary test on sparse graphs with nonuniform weights and native $R_{ZZ}$ interactions. At small scale, EP-Cost lowers native QPD overhead and mean TVD for all four sampled circuits. The parameter mechanism is clearest for Star and Erd\H{o}s--R\'enyi, where matched cut counts and implementation costs still yield baseline-to-EP-Cost overhead ratios of $2.41\times$ and $6.95\times$ because different $R_{ZZ}$ angles are cut. The reconstruction metrics are less uniform: Hellinger infidelity is effectively tied for the 3-regular instance, and Erd\H{o}s--R\'enyi shows a Max-Cut expectation-error reversal. Lower QPD burden therefore does not imply monotonic improvement in every application-specific observable.

In the controlled 24-qubit case, both strict 12/12 partitions cut four native $R_{ZZ}$ gates with equal implementation-cost contributions. EP-Cost reduces theoretical QPD overhead from 149.5 to 3.347. With 1,000 samples, it reduces generated subexperiments and mean runtime by factors of 4.15 and 6.21, respectively, while lowering weighted edge-marginal Hellinger infidelity. Exact-local results depend more strongly on bond dimension: against KL and spectral, win/loss/tie (W/L/T) counts change from 4/5/1 and 3/6/1 at $\chi=2$ to 9/1/0 and 8/1/1 at $\chi=8$.

At 50 qubits, weighted-QAOA is retained only as a reference-limited stress test because its higher-resource MPS references remain weakly converged. The complete comparisons and reference-adequacy statistics are reported in Appendix Table~\ref{tab:s5-50q-detailed} and Fig.~\ref{fig:s11-reference-adequacy}.

\FloatBarrier
\subsection{Cross-circuit finite-bond validation and optimizer robustness}
The held-out 24-qubit study tests the second-stage ranking signal on six previously unseen circuits before the larger benchmark and sensitivity analyses. Across the 18 instance--$\chi$ conditions, SACE has lower exact-local error than EP-only in 16 conditions and lower error than both KL and spectral in 11. The $\rmultichi$ ranking has positive Spearman association with exact local infidelity in all 18 comparisons, but the median correlations differ substantially: 0.930 for QFT and 0.418 for weighted-QAOA. The median selector-to-panel-oracle gap is 0.0093, with two comparisons above 0.10. The circuit-specific correlations reveal differences obscured by the pooled counts.

\begin{table}[!htb]
  \centering
  \small
  \renewcommand{\arraystretch}{1.08}
  \caption{Held-out 24-qubit validation of SACE. A ``win'' denotes lower isolated-local exact-vs-target-$\chi$ worst-block infidelity. The 18 comparisons arise from six held-out circuit instances evaluated at $\chi\in\{2,4,8\}$ and are repeated conditions within six circuits rather than 18 independent circuit draws.}
  \label{tab:heldout-24q-summary}
  \begingroup
  \setlength{\tabcolsep}{8pt}
  \begin{tabular*}{\linewidth}{@{\extracolsep{\fill}}lr@{}}
    \toprule
    Quantity & Result \\
    \midrule
    Held-out circuits & 6 (3 QFT-24 + 3 QAOA-24) \\
    Target $\chi$ values & 2, 4, 8 \\
    SACE beats EP-only & 16/18 \\
    SACE beats both KL and spectral & 11/18 \\
    Positive $\rmultichi$--$\localinf$ Spearman comparisons & 18/18 \\
    Median Spearman, QFT / QAOA & 0.930 / 0.418 \\
    Median selector--oracle gap & 0.0093 \\
    Gap $>0.10$ & 2/18 \\
    \bottomrule
  \end{tabular*}
  \endgroup
\end{table}

\begin{figure}[!htbp]
  \centering
  \includegraphics[width=\linewidth,keepaspectratio]{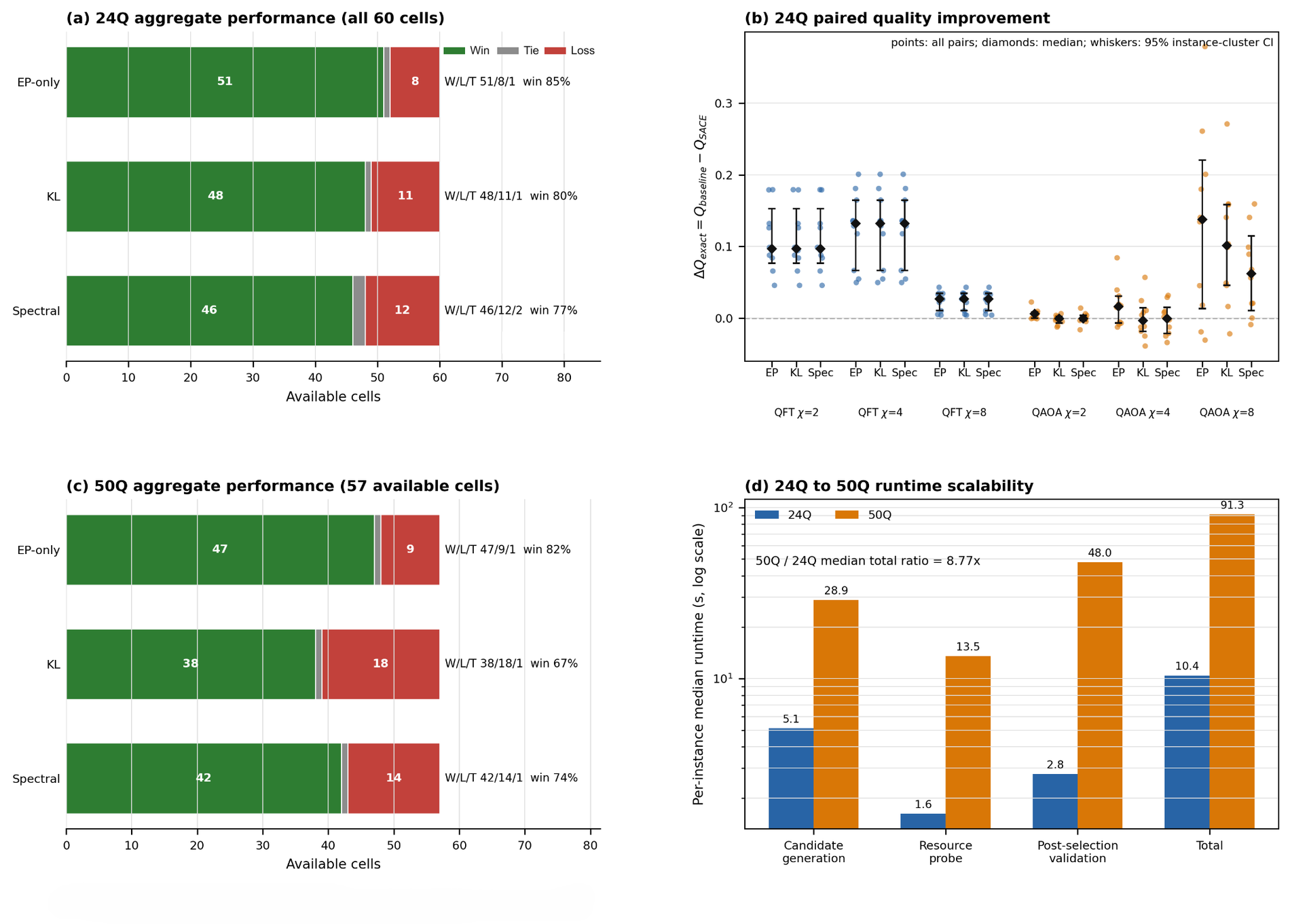}
  \caption{Aggregate 24/50-qubit benchmark overview. Panels (a) and (c) show aggregate win/tie/loss outcomes over the available 24- and 50-qubit comparisons. Panel (c) includes the reference-limited QAOA-50 cases; the primary 50-qubit interpretation uses QFT (Table~\ref{tab:large-scale-benchmark-summary}). Panel (b) shows paired 24-qubit improvements $\Delta Q_{\mathrm{exact}}=Q_{\mathrm{baseline}}-Q_{\mathrm{SACE}}$; points are paired comparisons, diamonds are medians, and whiskers are instance-cluster bootstrap intervals. Panel (d) reports median per-instance runtime by stage and width. The 24-qubit and 50-qubit validation quantities have different meanings and are not numerically pooled.}
  \Description[Aggregate 24/50-qubit benchmark]{Four panels summarize the larger-width benchmark. The first and third panels use stacked outcome bars for wins, ties, and losses at 24 and 50 qubits. The 50-qubit aggregate retains weighted-QAOA stress-test cases for completeness, while the primary 50-qubit quantitative interpretation is based on QFT because its higher-resource reference is better converged. The second panel shows paired 24-qubit quality improvements with median markers and bootstrap intervals. The fourth panel compares median runtime contributions from candidate generation, finite-bond probing, and post-selection validation.}
  \label{fig:large-scale-benchmark-overview}
\end{figure}

Table~\ref{tab:large-scale-benchmark-summary} summarizes the primary larger-width evidence: exact-local validation for QFT and weighted-QAOA at 24 qubits, and higher-resource MPS consistency for QFT at 50 qubits. Figure~\ref{fig:large-scale-benchmark-overview} provides the aggregate 24/50-qubit overview and runtime breakdown. Its 50-qubit aggregate retains the weighted-QAOA stress-test cases for completeness, but these cases are not used as primary 50-qubit accuracy evidence because their higher-resource references remain weakly converged.

\begin{table}[!htb]
\centering
\footnotesize
\caption{Primary larger-width benchmark summary. W/L/T denotes win/loss/tie relative to each baseline. The 24-qubit metric is exact-local MPS error for QFT and weighted-QAOA, whereas the 50-qubit row reports higher-resource MPS consistency for QFT. Here $n$ counts instance--$\chi$ conditions (ten circuit instances at three bond dimensions), not independent circuit draws.}
\label{tab:large-scale-benchmark-summary}
\begin{tabular*}{\linewidth}{@{\extracolsep{\fill}}llccccc@{}}
\toprule
Scale & Circuit & $n$ & vs EP-only & vs KL & vs Spec. & \shortstack{$A_{\mathrm{ref},\chi}$\\$\chi=2/4/8$} \\
\midrule
24Q & QFT  & 30 & 30/0/0 & 30/0/0 & 30/0/0 & -- \\
24Q & QAOA & 30 & 21/8/1 & 18/11/1 & 16/12/2 & -- \\
50Q & QFT  & 30 & 25/4/1 & 24/5/1 & 24/5/1 & 0.4274 / 0.1030 / 0.0167 \\
\bottomrule
\end{tabular*}
\end{table}

\noindent\begin{minipage}{\linewidth}
\footnotesize
At 24 qubits, validation uses isolated-local exact-vs-MPS infidelity $Q_{\mathrm{exact},\chi}$. The 50-qubit QFT row uses $Q_{\mathrm{highref},\chi}$ and must be interpreted together with the reference-adequacy diagnostic $A_{\mathrm{ref},\chi}$. The weighted-QAOA-50 results are excluded from this primary summary because their higher-resource references are weakly converged; they are reported in Appendix Table~\ref{tab:s5-50q-detailed}.
\end{minipage}

Top-32 contains a Top-64 exact-local co-optimum in 34/60 instance--$\chi$ conditions: 25/30 for QFT and 9/30 for weighted-QAOA. The two panel sizes select the same partition in 30/60 conditions. Expanding to Top-64 improves or preserves $\rmultichi$, while exact-local error improves in 18 conditions, is unchanged in 30, and worsens in 12. We retain $M=32$ as a fixed budget without assuming candidate-set saturation. Appendix Fig.~\ref{fig:topm-sensitivity-combined} and Table~\ref{tab:topm-sensitivity-summary} give the full results.

\FloatBarrier

The refinement benefit also persists across search procedures. Relative to each optimizer's own EP-only selection, SACE improves 23/30 conditions for GA, 29/30 for SA, and 24/30 for DPSO; all three optimizers improve all 15 QFT conditions, whereas the weighted-QAOA win counts are 8/15, 14/15, and 9/15. Different candidate sets can yield similar median local accuracy: GA and DPSO have median Top-32 Jaccard overlap 0.730 and identical selections in 22/30 conditions, while GA--SA and SA--DPSO have median overlap 0.008 and no identical selections. Median absolute validation-error differences are 0.030 for GA--SA and 0.025 for SA--DPSO, with larger differences in some instances. Full baseline comparisons and search-overlap diagnostics are reported in Appendix Table~\ref{tab:s6-optimizer-robustness}.

\begin{figure}[!htbp]
  \centering
  \includegraphics[width=\linewidth,keepaspectratio]{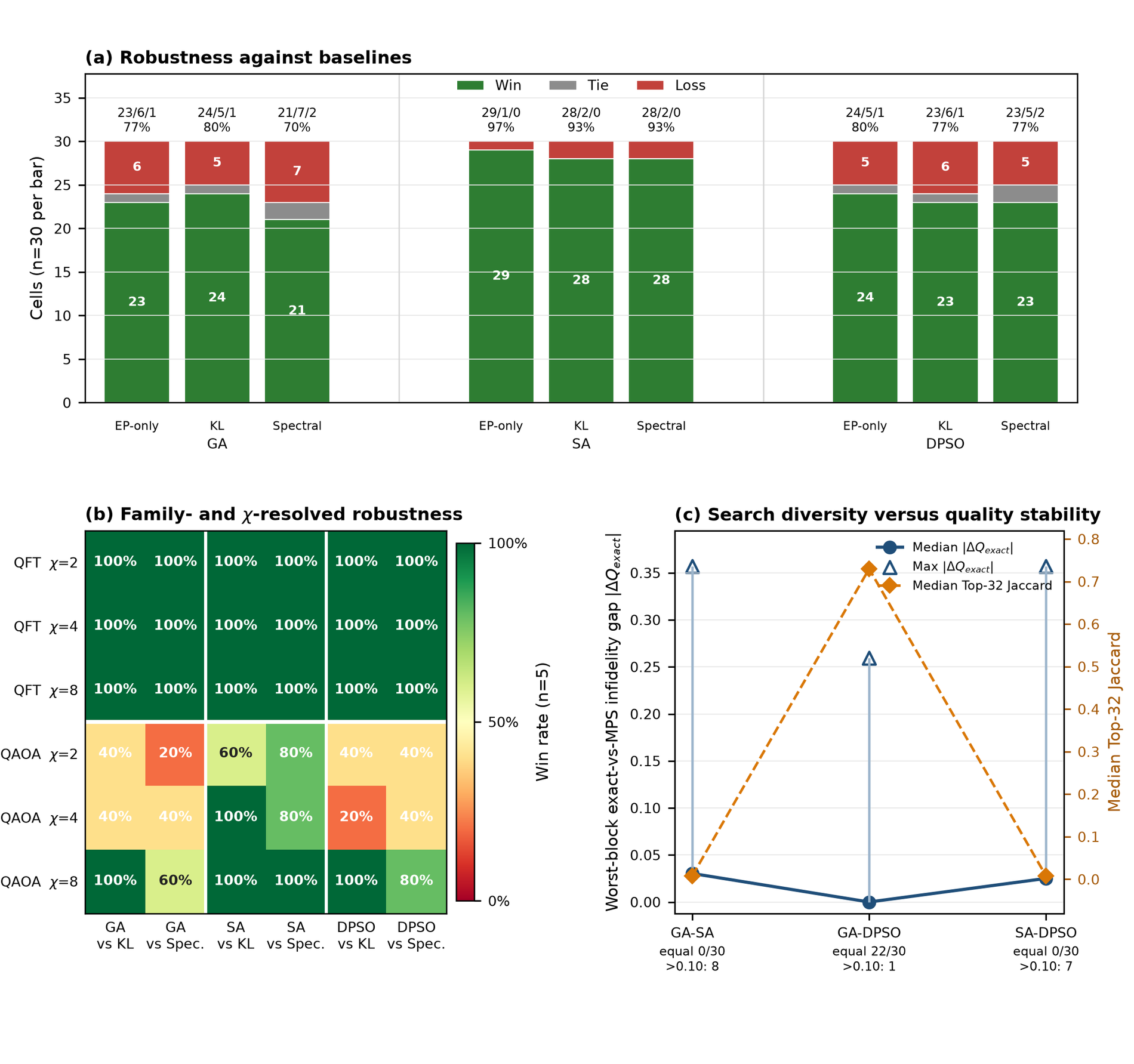}
  \caption{Optimizer robustness of SACE at 24 qubits under matched EP-Cost search budgets of three seeds with 2,000 evaluations each (6,000 per optimizer and circuit instance). (a) Aggregate SACE-vs-baseline win/tie/loss counts for GA, SA, and DPSO. (b) Win rates separated by circuit type and $\chi$ against KL and spectral baselines. (c) Search-diversity diagnostics versus exact-local quality stability for optimizer pairs. Search overlap is descriptive and is not used by SACE.}
  \Description[Optimizer robustness of SACE]{Three panels compare GA, SA, and DPSO. The first shows stacked win, tie, and loss bars; the second is a heat map of win rates separated by circuit type and bond dimension; the third compares Top-32 search overlap with exact-local quality gaps.}
  \label{fig:optimizer-robustness}
\end{figure}
\FloatBarrier

\section{Discussion}
The screening stage has both an analytical basis and empirical support. Its entanglement-capability term is the squared maximal product-input concurrence, which is monotonically related to native single-gate QPD overhead for CP and $R_{ZZ}$ gates. The GA-selected partition is QPD-overhead co-optimal in all 14 exhaustive small-scale spaces, with strong rank agreement between EP-Cost and native overhead. These findings support $\eta_g$ for screening the tested gate sets; they do not equate it with a QPD coefficient norm or establish the same relation for every two-qubit gate.

Finite-bond refinement adds local simulation information to boundary-based screening. On AQFT-12, lower isolated-subcircuit discrepancy accompanies lower local QPD-term error and better reconstruction in most low-bond conditions. With cut count and QPD term count fixed, SACE is best on all four metrics in four of six conditions, so the gains cannot be attributed solely to additional cuts or terms. The Top-$M$ analysis further distinguishes improvement in the finite-bond selection score from improvement in exact-local error. We retain $M=32$ as a fixed intermediate candidate-panel budget.

QFT gives the most consistent larger-width results: at 24 qubits, all 30 exact-local comparisons favor SACE against each of EP-only, KL, and spectral partitioning, and at 50 qubits reference convergence improves with $\chi$. Weighted-QAOA is more sensitive to bond dimension and search procedure and serves as a complementary stress test. Refinement benefits persist across GA, SA, and DPSO under matched search budgets. The 50-qubit weighted-QAOA results are retained in the appendix as a reference-convergence stress test.

SACE addresses partition selection within distributed quantum and quantum--HPC workflows. Validation spans sampled reconstruction at small scale, complete deterministic reconstruction on one 12-qubit AQFT topology, exact isolated-local states at 24 qubits, and finite-resource MPS references at 50 qubits. The larger-width results therefore concern local accuracy and consistency, not global reconstruction fidelity. Physical-QPU evaluation, inter-QPU communication, scheduling, and hardware-specific cutting costs remain outside the present scope. On physical QPUs, gate errors, decoherence, and readout errors may bias local estimates, while QPD coefficient weights can increase sampling variance. These effects motivate hardware-aware partition selection and experimental verification.

\section{Conclusion}

SACE addresses circuit-partition selection within distributed quantum and quantum--HPC workflows through two stages. EP-Cost screens boundaries using gate-level entanglement capability and native implementation cost; finite-bond refinement then selects the retained candidate that minimizes the worst-subcircuit discrepancy between MPS simulations at $\chi$ and $2\chi$. External constraints on cut count, QPD term count, and overhead define reconstruction feasibility. This design accounts for both boundary cost and local simulation behavior while keeping reconstruction-quality measurements separate from selection.

The experiments support both stages, with the most consistent results on QFT. EP-Cost-guided selections are co-optimal for native QPD overhead in all 14 exhaustive small-scale spaces. At matched cut counts, QFT-8 and QFT-9 achieve overhead reductions of $25.20\times$ and $16.89\times$, respectively, with improved sampled reconstruction. AQFT-12 extends the evaluation to complete deterministic reconstruction: among partitions with five cuts and 7,776 QPD terms, SACE achieves the lowest errors on all four reconstruction metrics in four of six low-bond conditions. At 24 qubits, all 30 QFT conditions favor SACE against each of EP-only, KL, and spectral partitioning. The 50-qubit QFT comparisons remain favorable as reference convergence improves. Refinement benefits persist across GA, SA, and DPSO, while weighted-QAOA shows greater sensitivity to bond dimension and search procedure.

Together, these findings support SACE as a resource-aware component for circuit-partition selection, combining reconstruction-aware boundary screening with finite-bond refinement. The consistent QFT improvements and the gains obtained across GA, SA, and DPSO demonstrate its potential for distributed quantum and quantum--HPC workflows. Future work will extend complete reconstruction to larger circuits, strengthen scalable validation for challenging circuit classes, and incorporate hardware-specific cutting costs, noise, communication, and scheduling into multi-QPU execution.

\begin{acks}
This work is supported by the National Key Research and Development Program of China (2024YFB4504103). This work is also supported by Jiangsu Province Engineering Research Center of IntelliSense Technology and System.
\end{acks}

\section*{Competing Interests}
The authors declare no competing interests.




\appendix
\setcounter{figure}{0}
\renewcommand{\thefigure}{S\arabic{figure}}
\renewcommand{\theHfigure}{S\arabic{figure}}
\setcounter{table}{0}
\renewcommand{\thetable}{S\arabic{table}}
\renewcommand{\theHtable}{S\arabic{table}}
\section*{Appendix}
The appendices report candidate-space and QPD-oracle analysis, matched-resource AQFT controls, controlled-QAOA reconstruction, candidate-panel sensitivity, finite-bond diagnostics, and detailed benchmark results.

\section{Small-Scale Candidate-Space and Sampled-QPD Results}
\label{app:small-scale-candidate-space}

\begin{figure}[!htbp]
  \centering
  \includegraphics[width=0.68\linewidth,keepaspectratio]{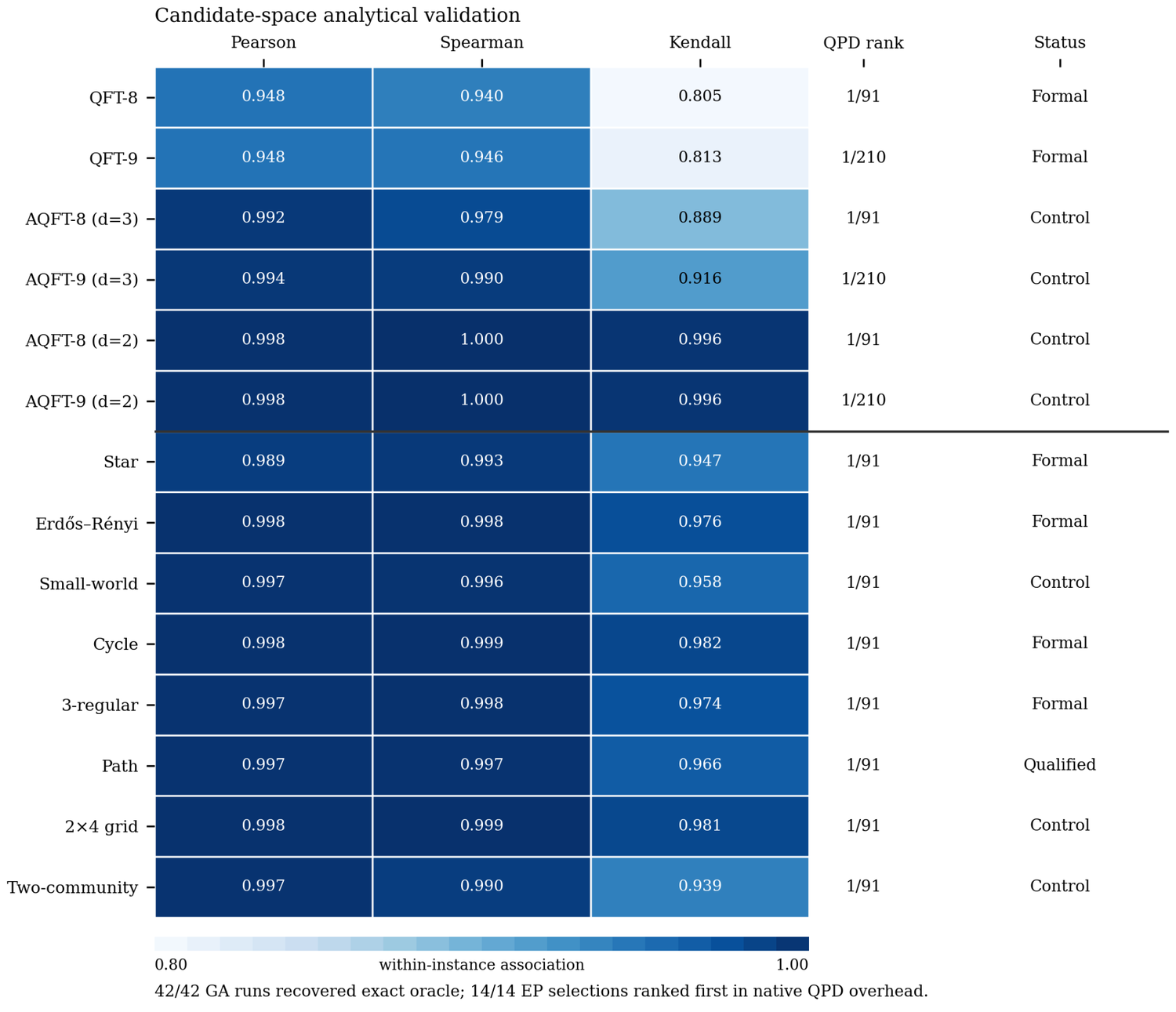}
  \caption{Candidate-space analytical validation and QPD-overhead oracle alignment. Pearson, Spearman, and Kendall $\tau_b$ quantify the within-instance association between EP-Cost and $\log_{10}$ native QPD overhead for all 14 candidate circuits. The QPD-rank column gives the tie-aware native-overhead rank of the EP-Cost-selected partition; rank 1 denotes membership in the QPD-overhead co-optimal set. Statistics are computed within each candidate space and are not pooled across circuit types.}
  \Description[Candidate-space analytical validation]{A heat-map table lists 14 QFT, approximate-QFT, and weighted-QAOA candidate circuits. Columns report Pearson, Spearman, and Kendall associations, native-QPD-overhead rank, and each circuit's evaluation role.}
  \label{fig:small-scale-candidate-space}
\end{figure}

\paragraph{Candidate-space validation.}
Figure~\ref{fig:small-scale-candidate-space} reports within-circuit associations and native-QPD-overhead ranks for all 14 exhaustive candidate spaces. All 42 GA searches recover an EP-Cost co-optimum. At the circuit level, every GA-selected partition belongs to the global $\Omega_{\mathrm{QPD}}$ co-optimal set, and the exact EP-Cost optimum intersects that set in all 14 spaces. The GA-to-QPD-oracle overhead ratio is therefore 1 throughout the tested spaces.

\begin{figure}[!htbp]
  \centering
  \includegraphics[width=0.84\linewidth,keepaspectratio]{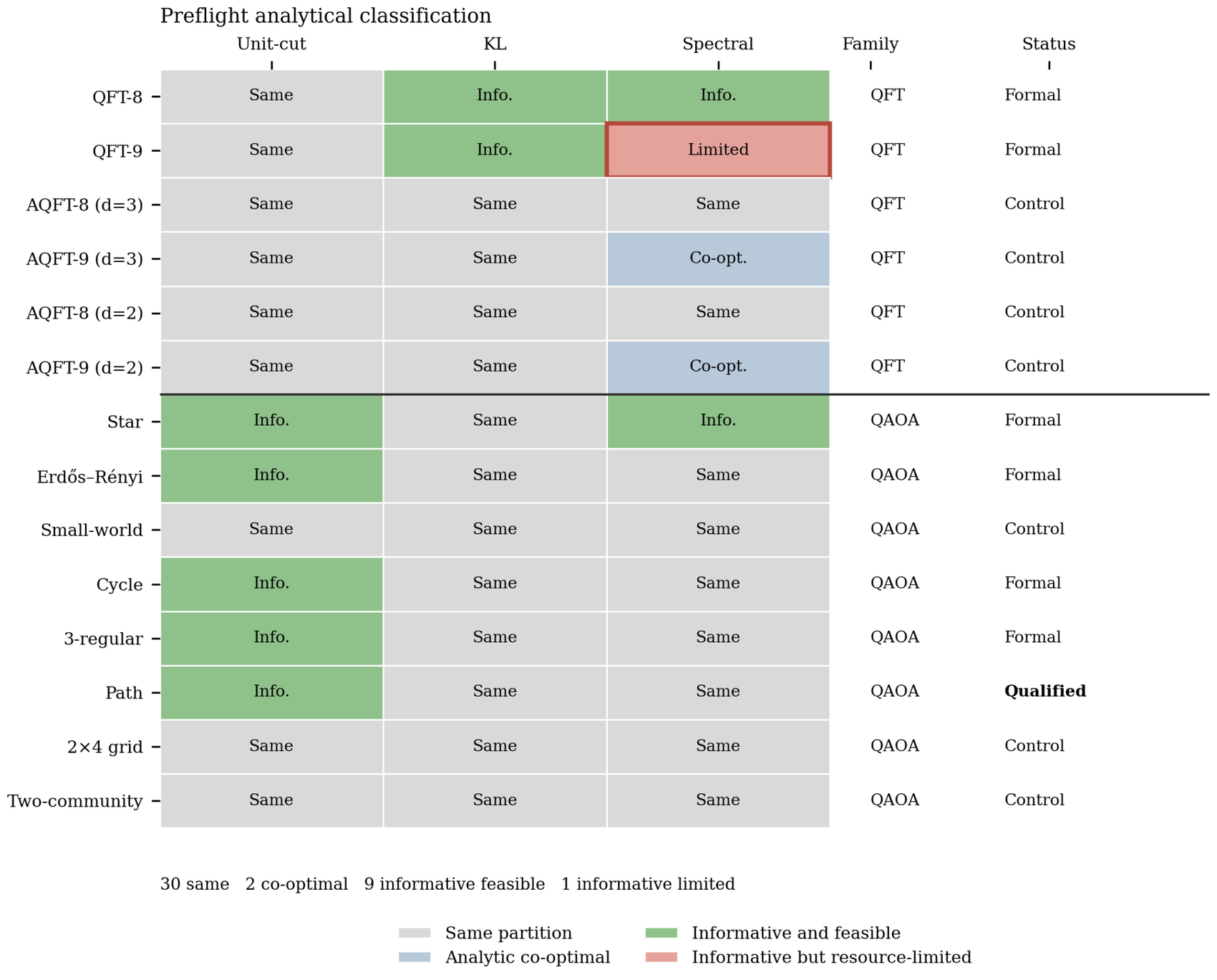}
  \caption{Pre-sampling candidate classification across the 14 benchmark circuits. ``Same'' denotes an identical canonical partition, ``Co-opt.'' a distinct analytically co-optimal partition, ``Info.'' an informative sampled-QPD-feasible comparison, and ``Limited'' an informative comparison exceeding the QPD resource limits. Classification is determined before reconstruction metrics are evaluated.}
  \Description[Pre-sampling candidate classification]{A matrix classifies each of 14 circuits against unit-cut, KL, and spectral baselines as identical, analytically co-optimal, informative and feasible, or informative but resource-limited. Additional columns identify circuit type and evaluation role.}
  \label{fig:small-scale-selection-audit}
\end{figure}

\paragraph{Pre-sampling comparison classification.}
Figure~\ref{fig:small-scale-selection-audit} classifies each candidate--baseline pair before sampled-QPD reconstruction is evaluated. Across the 14 circuits, 30 comparisons produce the same canonical partition, two produce distinct analytically co-optimal partitions, nine are informative and sampled-QPD-feasible, and one is informative but exceeds the declared reconstruction limits. This classification is determined without using TVD, Hellinger, or observable-error outcomes.

\paragraph{Detailed sampled-QPD metrics.}
Tables~\ref{tab:small-scale-full-sampled-metrics} and~\ref{tab:small-scale-execution-metrics} report reconstruction metrics, sampling diagnostics, and execution workload. Runtime depends on the execution environment.

\begin{table}[H]
\centering
\scriptsize
\caption{Full sampled-QPD metrics for the six benchmark circuits. QFT repeats comprise three input states crossed with three sampling seeds; QAOA repeats comprise three sampling seeds. Values are mean $\pm$ sample standard deviation. Runtime is environment-dependent, and repeat counts are not independent circuit instances.}
\label{tab:small-scale-full-sampled-metrics}
\begingroup
\setlength{\tabcolsep}{3.0pt}
\renewcommand{\arraystretch}{1.05}
\textbf{(a) Reconstruction and sampling diagnostics}\par\vspace{1mm}
\resizebox{\linewidth}{!}{%
\begin{tabular}{llcccccc}
\toprule
Circuit & Method & $n$ & Native $\Omega_{\mathrm{QPD}}$ & TVD & Hellinger fidelity & $I_H$ & Negative mass \\
\midrule
QFT-8 & EP-Cost & 9 & 2.67e+03 & .440 $\pm$ .153 & .598 $\pm$ .184 & .402 $\pm$ .184 & .295 $\pm$ .197 \\
QFT-8 & Primary baseline & 9 & 6.72e+04 & .565 $\pm$ .129 & .442 $\pm$ .153 & .558 $\pm$ .153 & 2.319 $\pm$ 1.036 \\
QFT-9 & EP-Cost & 9 & 3.16e+03 & .451 $\pm$ .123 & .581 $\pm$ .145 & .419 $\pm$ .145 & .394 $\pm$ .235 \\
QFT-9 & Primary baseline & 9 & 5.33e+04 & .558 $\pm$ .034 & .437 $\pm$ .046 & .563 $\pm$ .046 & 2.880 $\pm$ .598 \\
Star & EP-Cost & 3 & 61.4 & .080 $\pm$ .009 & .972 $\pm$ .003 & .028 $\pm$ .003 & .005 $\pm$ .001 \\
Star & Primary baseline & 3 & 148 & .103 $\pm$ .012 & .951 $\pm$ .008 & .049 $\pm$ .008 & .018 $\pm$ .003 \\
Erd\H{o}s--R\'enyi & EP-Cost & 3 & 51.3 & .077 $\pm$ .005 & .981 $\pm$ .002 & .019 $\pm$ .002 & .003 $\pm$ .002 \\
Erd\H{o}s--R\'enyi & Primary baseline & 3 & 356 & .094 $\pm$ .006 & .962 $\pm$ .001 & .038 $\pm$ .001 & .027 $\pm$ .001 \\
Cycle & EP-Cost & 3 & 20.2 & .081 $\pm$ .004 & .980 $\pm$ .003 & .020 $\pm$ .003 & .005 $\pm$ .000 \\
Cycle & Primary baseline & 3 & 34.2 & .089 $\pm$ .001 & .975 $\pm$ .001 & .025 $\pm$ .001 & .007 $\pm$ .002 \\
3-regular & EP-Cost & 3 & 128 & .097 $\pm$ .003 & .966 $\pm$ .002 & .034 $\pm$ .002 & .014 $\pm$ .003 \\
3-regular & Primary baseline & 3 & 163 & .098 $\pm$ .006 & .966 $\pm$ .003 & .034 $\pm$ .003 & .013 $\pm$ .004 \\
\bottomrule
\end{tabular}%
}
\endgroup
\end{table}

\begin{table}[H]
\centering
\scriptsize
\caption{Execution workload and QAOA observable errors for the six sampled-QPD benchmark circuits. Values are mean $\pm$ sample standard deviation. Runtime is reported as an environment-dependent diagnostic rather than a portable complexity measure.}
\label{tab:small-scale-execution-metrics}
\begingroup
\setlength{\tabcolsep}{3.0pt}
\renewcommand{\arraystretch}{1.05}
\resizebox{\linewidth}{!}{%
\begin{tabular}{llcccc}
\toprule
Circuit & Method & Generated subexperiments & Runtime (s) & Max-Cut abs. error & Max-Cut rel. error \\
\midrule
QFT-8 & EP-Cost & 1810.7 $\pm$ 4.0 & 93.119 $\pm$ 3.185 & -- & -- \\
QFT-8 & Primary baseline & 1968.7 $\pm$ 5.6 & 103.894 $\pm$ 3.617 & -- & -- \\
QFT-9 & EP-Cost & 1844.7 $\pm$ 7.0 & 128.461 $\pm$ 8.992 & -- & -- \\
QFT-9 & Primary baseline & 1974.7 $\pm$ 3.6 & 144.256 $\pm$ 2.762 & -- & -- \\
Star & EP-Cost & 349.3 $\pm$ 1.2 & 7.434 $\pm$ .134 & .034 $\pm$ .025 & .013 $\pm$ .009 \\
Star & Primary baseline & 384.7 $\pm$ 4.6 & 7.869 $\pm$ .141 & .045 $\pm$ .020 & .017 $\pm$ .007 \\
Erd\H{o}s--R\'enyi & EP-Cost & 340.0 $\pm$ 4.0 & 8.364 $\pm$ .282 & .043 $\pm$ .051 & .011 $\pm$ .013 \\
Erd\H{o}s--R\'enyi & Primary baseline & 432.0 $\pm$ .0 & 10.654 $\pm$ .168 & .021 $\pm$ .020 & .005 $\pm$ .005 \\
Cycle & EP-Cost & 72.0 $\pm$ .0 & 1.791 $\pm$ .007 & .040 $\pm$ .014 & .011 $\pm$ .004 \\
Cycle & Primary baseline & 72.0 $\pm$ .0 & 1.583 $\pm$ .074 & .074 $\pm$ .065 & .021 $\pm$ .018 \\
3-regular & EP-Cost & 382.0 $\pm$ 4.0 & 12.177 $\pm$ .295 & .058 $\pm$ .030 & .014 $\pm$ .007 \\
3-regular & Primary baseline & 398.0 $\pm$ .0 & 11.969 $\pm$ .003 & .092 $\pm$ .026 & .022 $\pm$ .006 \\
\bottomrule
\end{tabular}%
}
\endgroup
\end{table}

\FloatBarrier
\section{Matched-Resource AQFT-12 Controls}
\label{app:aqft-matched-resource}
The main AQFT-12 comparison uses different reconstruction resources for SACE and the shared KL/Spectral control. To isolate partition choice, the matched-resource analysis evaluates all three feasible five-cut partitions in each of the six low-bond cells. Every candidate in this comparison has $K=5$ and $T_{\mathrm{QPD}}=7776$; $\Gamma$ and $\Omega_{\mathrm{QPD}}$ remain partition dependent.

\begin{figure}[!htbp]
  \centering
  \includegraphics[width=0.88\linewidth,keepaspectratio]{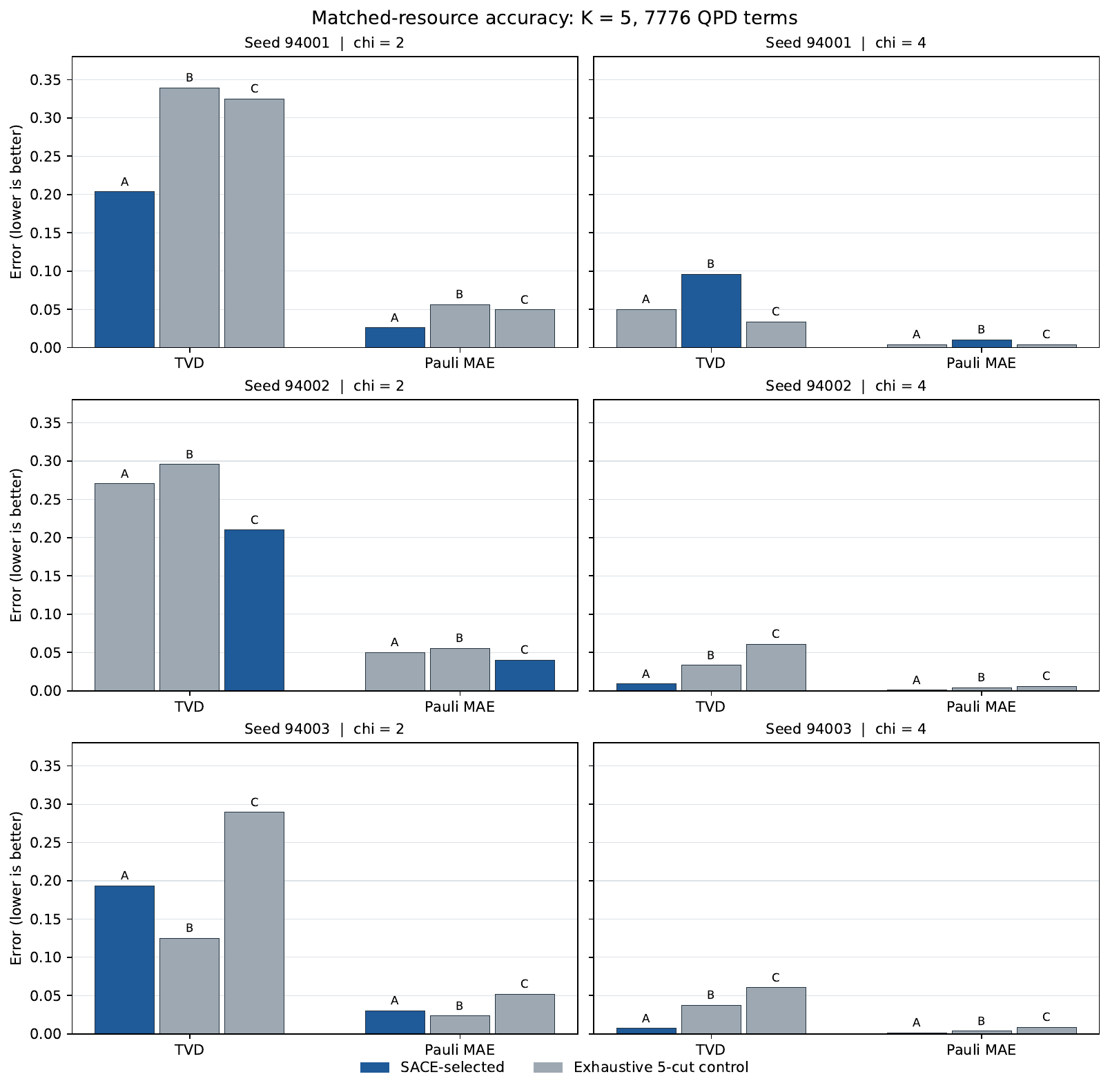}
  \caption{Matched-resource deterministic-QPD accuracy for AQFT-12. Each panel compares the three feasible five-cut partitions at a fixed input state and bond dimension. All candidates use five CP cuts and 7,776 deterministic-QPD tensor terms. Blue bars mark the partition selected by SACE; gray bars are the other exhaustive five-cut controls. Lower TVD and Pauli-observable MAE are better.}
  \Description[Matched-resource deterministic-QPD accuracy]{Six panels compare three five-cut AQFT partitions at three input states and bond dimensions 2 and 4. Each panel shows TVD and Pauli-observable mean absolute error. The SACE-selected partition is highlighted and the other two exhaustive matched-resource partitions are shown as controls.}
  \label{fig:aqft-matched-resource-accuracy}
\end{figure}

\begin{table}[!htbp]
\centering
\footnotesize
\setlength{\tabcolsep}{5pt}
\caption{Matched-resource AQFT-12 summary. For each metric, the SACE-selected partition is compared with the other two feasible five-cut partitions over six input-state--$\chi$ cells. $\Delta$ is SACE error minus the best alternative five-cut error; negative values favor SACE.}
\label{tab:aqft-matched-resource-summary}
\begin{tabular*}{\linewidth}{@{\extracolsep{\fill}}lccccc@{}}
\toprule
Metric & Best & Not best & Median rank & Mean $\Delta$ & Median $\Delta$ \\
\midrule
TVD & 4/6 & 2/6 & 1 & $-0.017551$ & $-0.027337$ \\
Hellinger infidelity & 4/6 & 2/6 & 1 & $-0.005311$ & $-0.006528$ \\
Pauli MAE & 4/6 & 2/6 & 1 & $-0.004180$ & $-0.002771$ \\
Raw $L_1$ & 4/6 & 2/6 & 1 & $-0.020723$ & $-0.054833$ \\
\bottomrule
\end{tabular*}
\end{table}

For all four metrics, the SACE rank distribution is 4/1/1 across ranks 1/2/3, giving a mean rank of 1.5. In $\Omega_{\mathrm{QPD}}$--TVD and $\Omega_{\mathrm{QPD}}$--Pauli-MAE comparisons, the SACE-selected point is Pareto efficient in four of six cells. The shared KL/Spectral partition is Pareto efficient in all six because its three-cut construction has a much lower reconstruction-cost coordinate. The matched-resource result therefore supports a partition-selection contribution in most tested cells without implying uniform optimality.

\FloatBarrier
\section{Controlled 24-Qubit QAOA Evaluation}
\label{app:qaoa-controlled}
\begin{figure}[!htbp]
  \centering
  \includegraphics[width=1.0\linewidth,keepaspectratio]{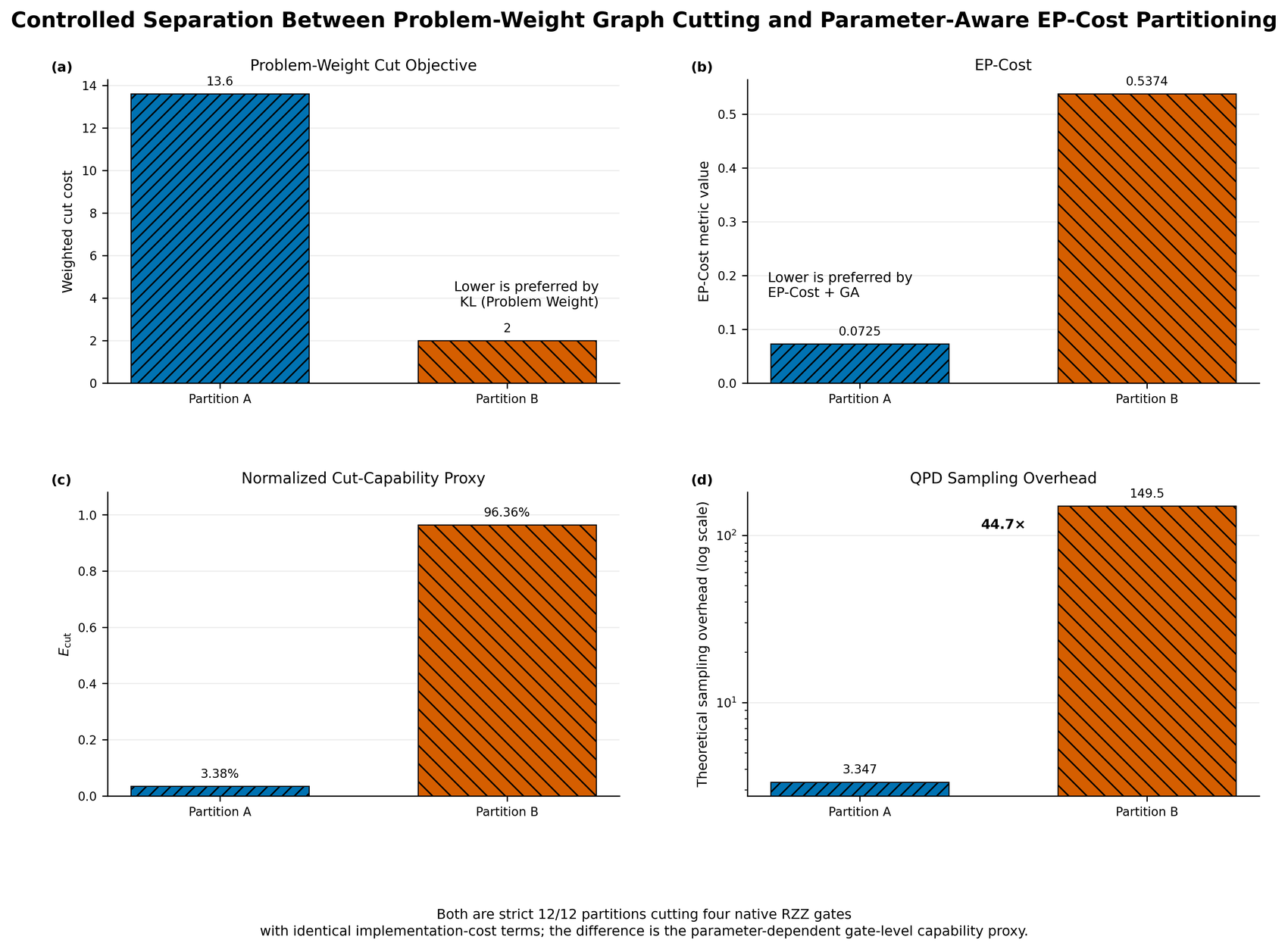}
  \caption{Controlled comparison of problem-weight graph partitioning and EP-Cost. Both candidates are strict 12/12 partitions with four native $R_{ZZ}$ cuts and identical implementation-cost contributions.}
  \Description[Controlled QAOA objective separation]{A four-panel bar-chart figure compares two equal-size and equal-cut-count QAOA partitions in terms of problem-weight cut objective, EP-Cost, normalized entanglement-capability loss, and theoretical QPD overhead.}
  \label{fig:qpd-controlled-separation}
\end{figure}

\begin{figure}[!htbp]
  \centering
  \includegraphics[width=0.9\linewidth,keepaspectratio]{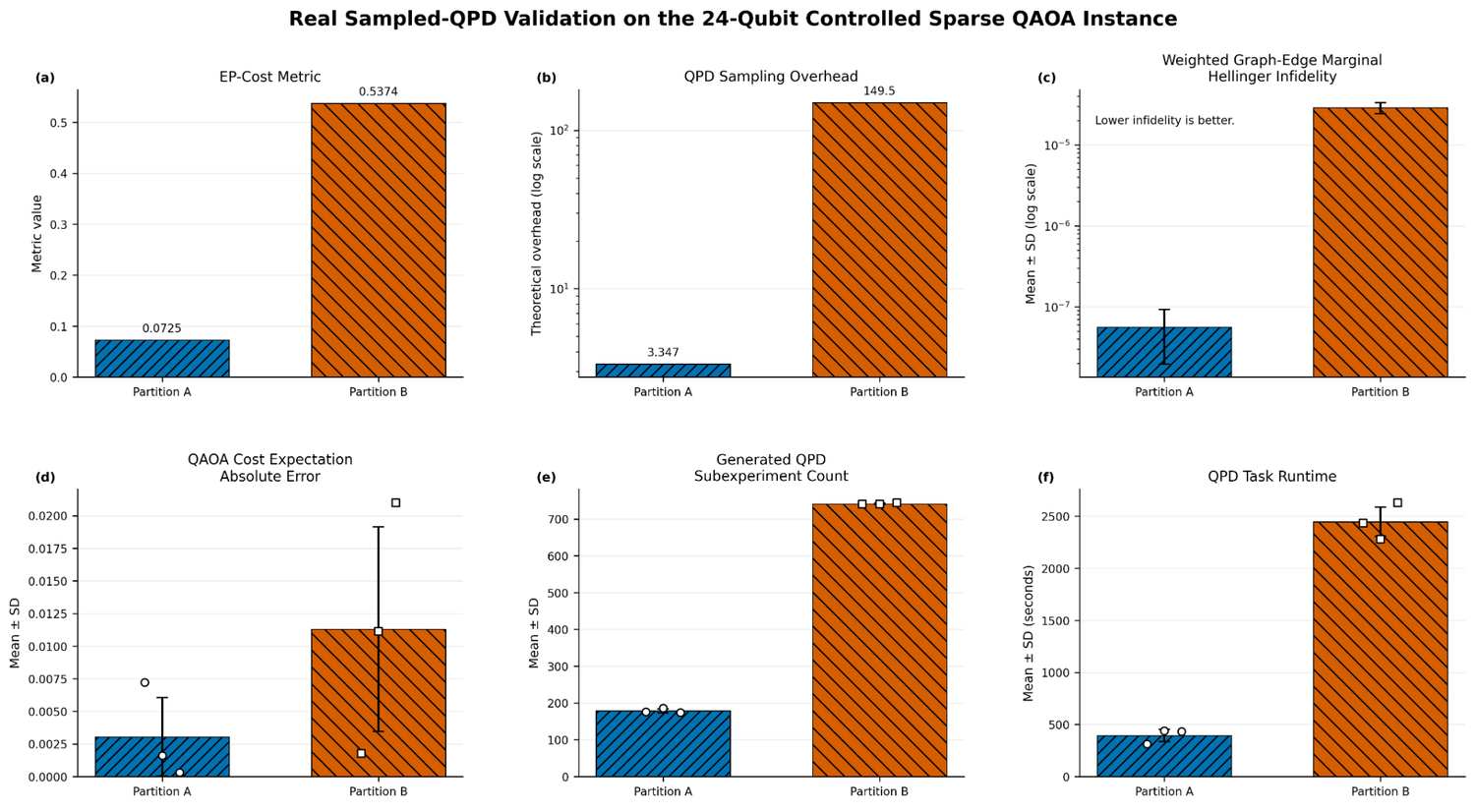}
  \caption{Controlled sampled-QPD evaluation of the two physical QAOA partitions.  The panels report gate-cut score, theoretical QPD overhead, weighted edge-marginal Hellinger infidelity, QAOA cost-expectation error, generated subexperiments, and end-to-end runtime at a 1,000-sample budget over three paired seeds.}
  \Description[Overview of the controlled QAOA sampled-QPD experiment]{A six-panel figure compares two physical QAOA partitions using gate-score, reconstruction-quality, and resource measurements.}
  \label{fig:qpd-real-overview}
\end{figure}

\begin{figure}[!htbp]
  \centering
  \includegraphics[width=0.9\linewidth,keepaspectratio]{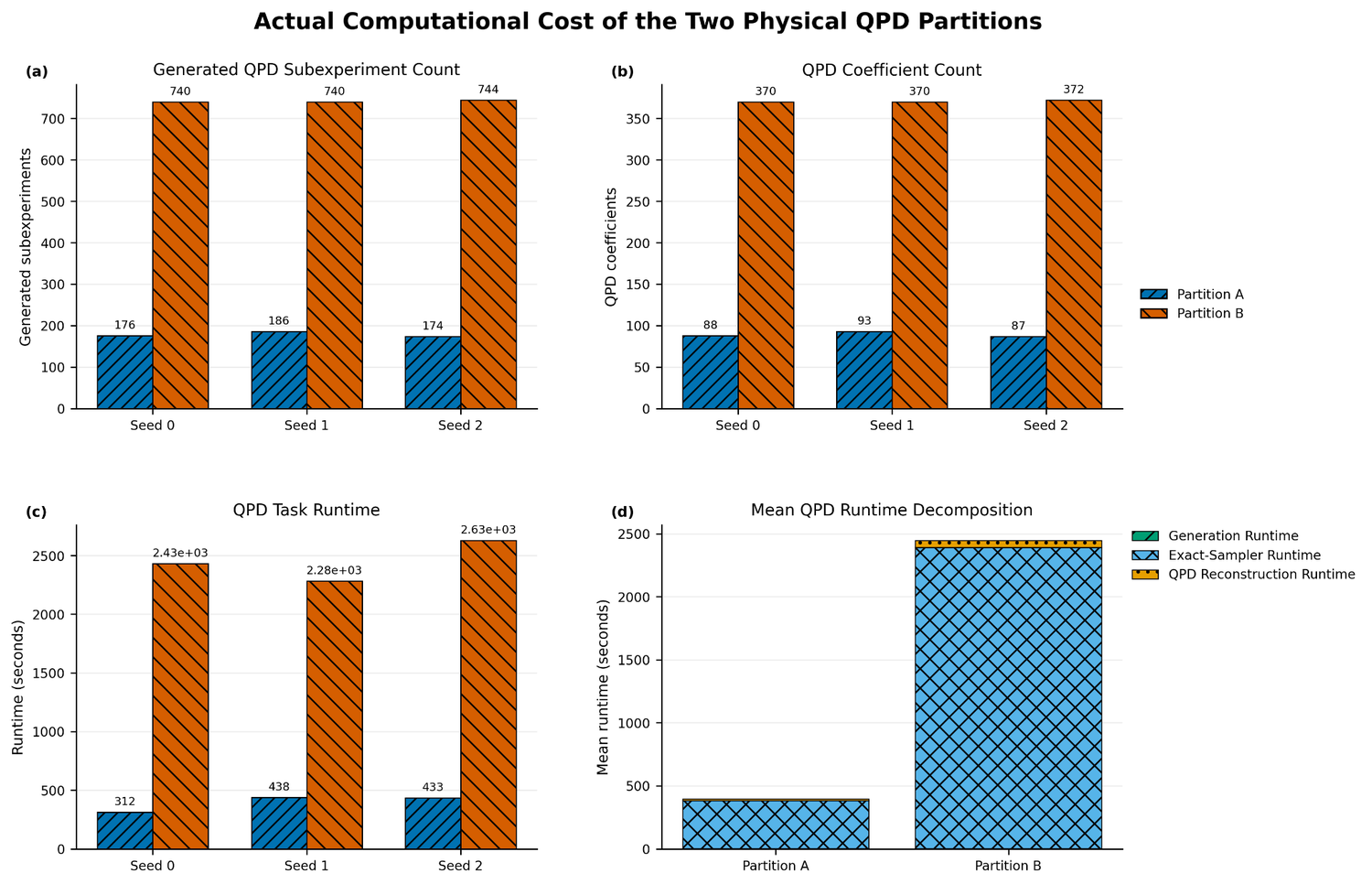}
  \caption{Measured computational cost of the two QPD partitions. Panels report generated subexperiment count, sampled coefficient-record count, end-to-end task runtime for each paired seed, and mean runtime decomposition. The sampled coefficient-record count is sample dependent and differs from the fixed exact decomposition size.}
  \Description[Measured QPD computational cost]{A four-panel figure compares generated QPD subexperiments, sampled coefficient-record counts, task runtimes, and mean runtime decomposition for two partitions across three paired seeds.}
  \label{fig:qpd-actual-cost}
\end{figure}

The controlled 24-qubit weighted-QAOA comparison fixes block size, cut count, and implementation cost while varying the parameters of the four cut $R_{ZZ}$ gates. Partition choices use only objective and feasibility information. Figures~\ref{fig:qpd-controlled-separation}--\ref{fig:qpd-actual-cost} report analytical costs, sampled reconstruction, and runtime.

Across three paired seeds, EP-Cost reduces generated subexperiments and mean wall-clock time by factors of 4.15 and 6.21, respectively. These reductions are smaller than the analytical overhead ratio. The sample-dependent coefficient-record count differs from the fixed exact tensor-product term count.

\FloatBarrier
\section{Finite-Bond Validation and Candidate-Panel Sensitivity}
\label{app:large-scale-validation}
The candidate-panel sensitivity analysis reuses the same twenty 24-qubit GA archives as the large-scale benchmark and forms nested Top-$M$ panels with $M\in\{8,16,32,64\}$. Top-32 results are unchanged from the primary evaluation; ranks 33--64 extend the same ordered archives. Table~\ref{tab:topm-sensitivity-summary} summarizes the effect of panel size relative to Top-64.

\begin{table}[!htbp]
\centering
\footnotesize
\setlength{\tabcolsep}{4pt}
\caption{Sensitivity to the EP-Cost-ranked candidate-panel size over 60 24-qubit instance--$\chi$ cells. $\Delta R$ and $\Delta Q$ denote the selected Top-$M$ value minus the selected Top-64 value for $\rmultichi$ and $Q_{\mathrm{exact},\chi}$, respectively. The candidate-evaluation ratio is determined by panel size and is not a wall-clock runtime ratio.}
\label{tab:topm-sensitivity-summary}
\begin{tabular*}{\linewidth}{@{\extracolsep{\fill}}lccccc@{}}
\toprule
$M$ & Candidate evals / $M{=}32$ & Same as $M{=}64$ & $R_{64}$ coverage & $Q_{64}$ coverage & Median $\Delta Q$ \\
\midrule
8  & 0.25 & 8/60  & 8/60  & 6/60  & 0.008179 \\
16 & 0.50 & 14/60 & 14/60 & 12/60 & 0.003828 \\
32 & 1.00 & 30/60 & 30/60 & 34/60 & 0 \\
64 & 2.00 & 60/60 & 60/60 & 60/60 & 0 \\
\bottomrule
\end{tabular*}
\end{table}

Top-32 covers a Top-64 exact-local co-optimum in 25/30 QFT cells and 9/30 weighted-QAOA cells. From $M=32$ to $M=64$, exact-local error improves in 18 cells, is unchanged in 30, and worsens in 12, even though the optimized $\rmultichi$ value never worsens. This difference is consistent with using $\rmultichi$ as a selection proxy rather than an exact-local error objective.

\begin{figure}[!htbp]
  \centering
  \includegraphics[width=0.82\linewidth,keepaspectratio]{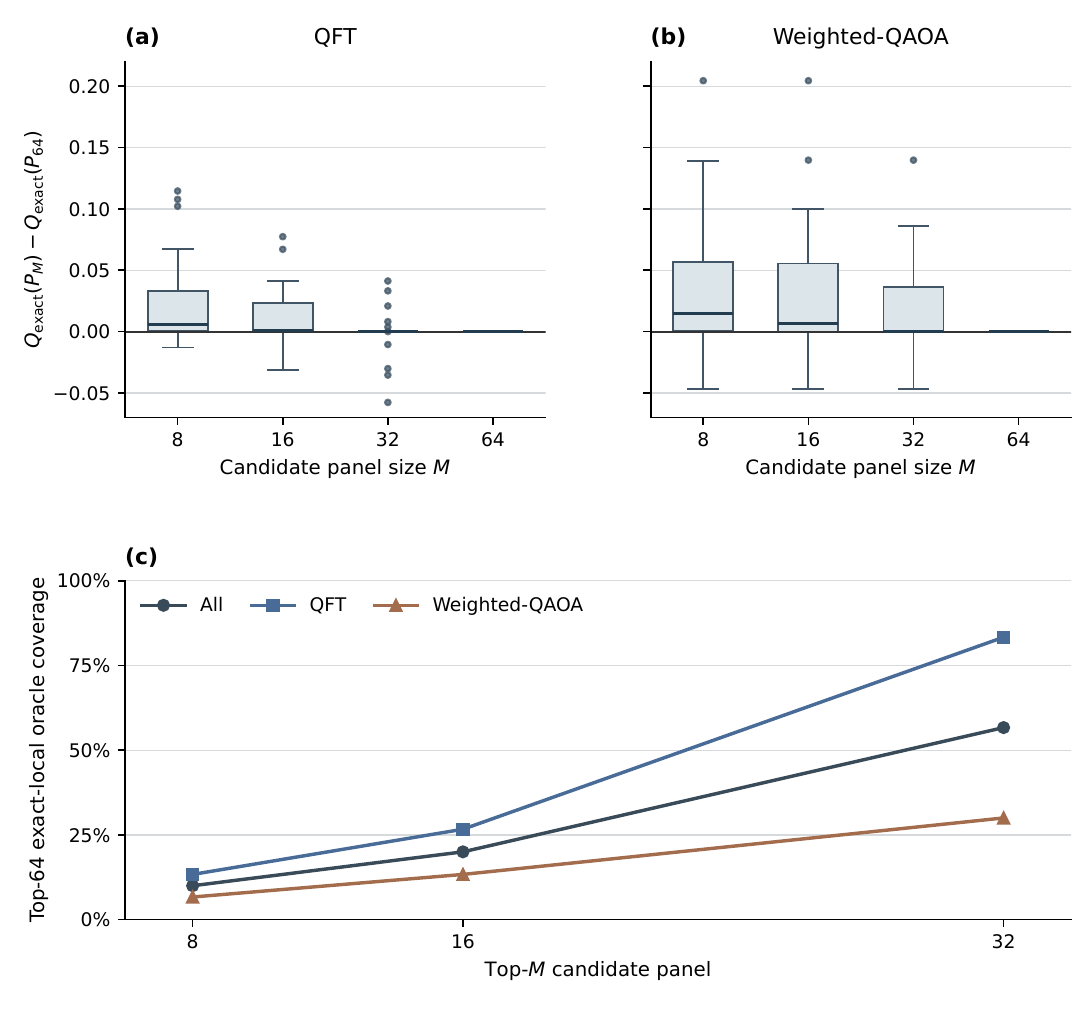}
  \caption{Sensitivity to the EP-Cost-ranked candidate-panel size at 24 qubits. Panels (a) and (b) show $Q_{\mathrm{exact},\chi}(P_M)-Q_{\mathrm{exact},\chi}(P_{64})$ for QFT and weighted-QAOA over $M\in\{8,16,32,64\}$; zero indicates the same exact-local error as the Top-64 SACE selection. Panel (c) shows the fraction of Top-64 exact-local co-optima already contained in Top-$M$ for all cells, QFT, and weighted-QAOA. The oracle is defined only within the evaluated Top-64 panel.}
  \Description[Top-M candidate-panel sensitivity]{A combined three-panel figure. Two box plots show exact-local error differences relative to the Top-64 SACE selection for QFT and QAOA as the candidate-panel size increases. A third panel shows Top-64 exact-local oracle coverage for all cells, QFT, and QAOA at panel sizes 8, 16, and 32.}
  \label{fig:topm-sensitivity-combined}
\end{figure}

\begin{figure}[!htbp]
  \centering
  \includegraphics[width=0.9\linewidth,keepaspectratio]{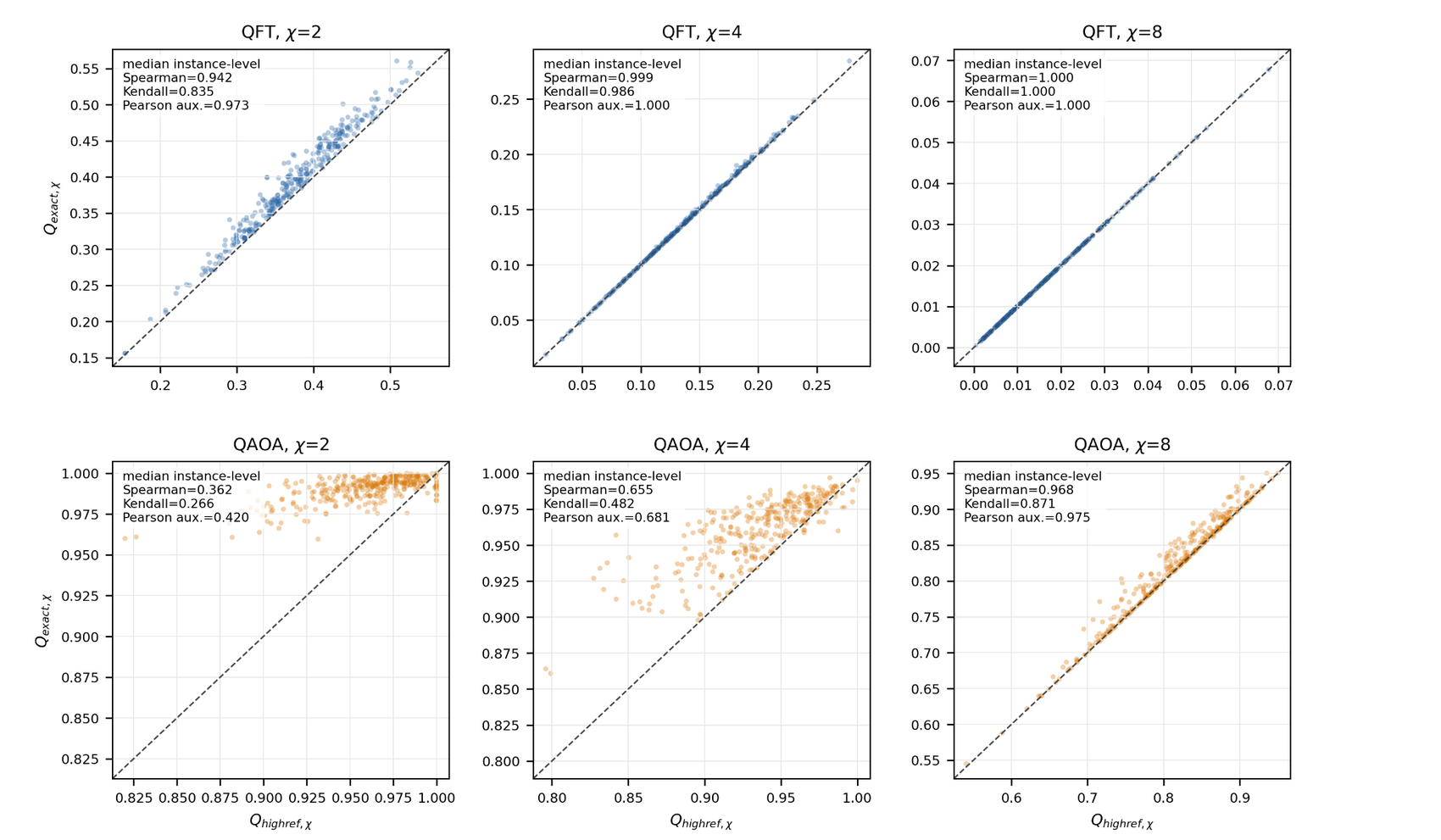}
  \caption{24-qubit relation between higher-resource MPS consistency and isolated-local exact-vs-MPS error over fixed Top-32 candidate panels. Spearman and Kendall values are medians of within-instance statistics. Associations are consistently high for QFT and increase with $\chi$ for QAOA.}
  \Description[24-qubit validation bridge]{Six scatter plots compare higher-resource MPS consistency on the horizontal axis with exact local MPS error on the vertical axis for QFT and QAOA at bond dimensions 2, 4, and 8. Dashed diagonals mark equality.}
  \label{fig:s10-validation-bridge}
\end{figure}

At 24 qubits, Fig.~\ref{fig:s10-validation-bridge} compares higher-resource MPS consistency with exact-local error. Correlations are computed within each instance and summarized by their medians; Top-32 panel points are not pooled as independent observations. Spearman and Kendall coefficients summarize rank association, with Pearson correlation reported descriptively. Agreement is strong for QFT at all tested bond dimensions and increases with $\chi$ for QAOA.

\begin{figure}[!htbp]
  \centering
  \includegraphics[width=0.82\linewidth,keepaspectratio]{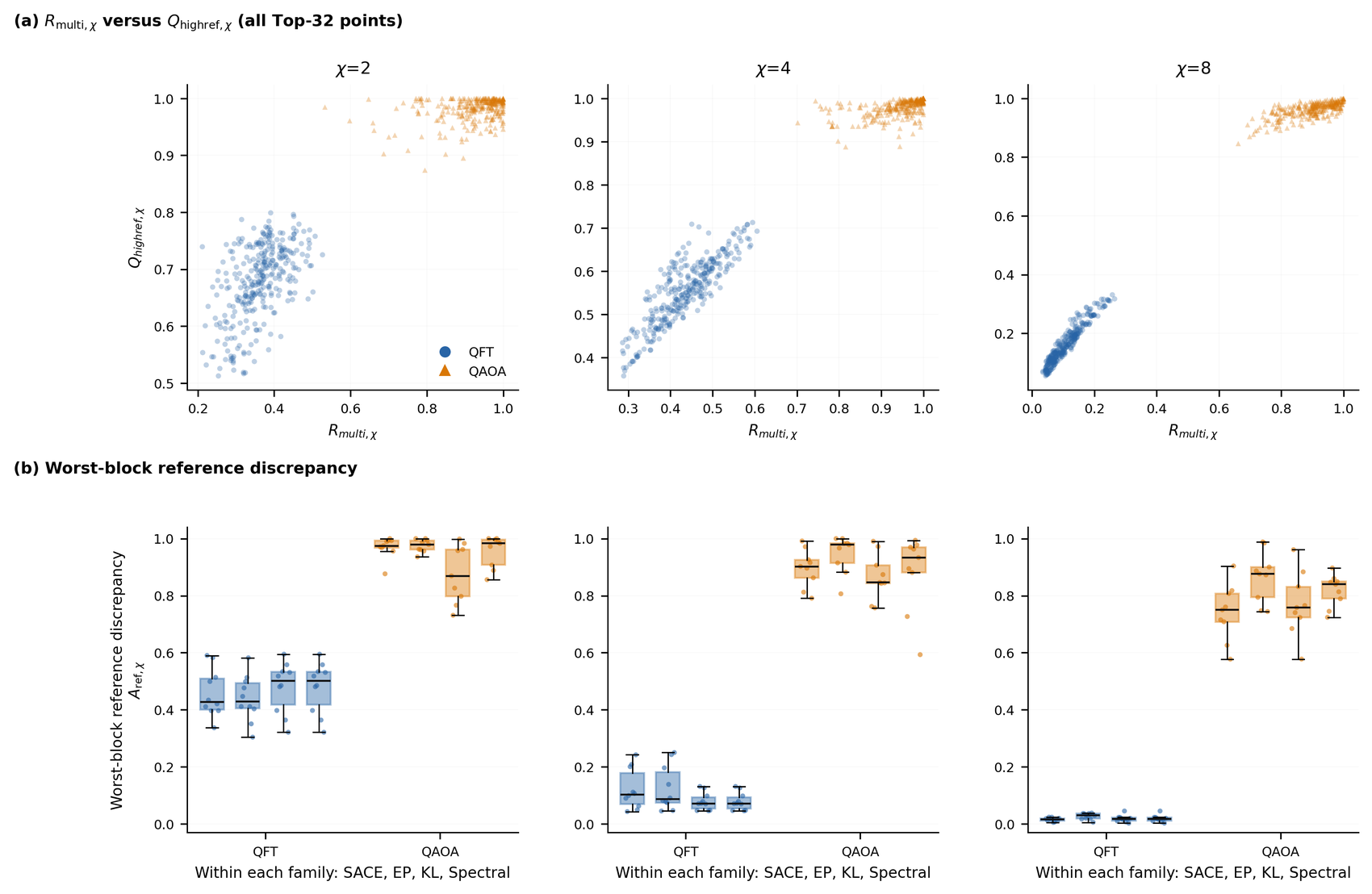}
  \caption{50-qubit higher-resource validation and reference adequacy. The upper row relates $\rmultichi$ to $Q_{\mathrm{highref},\chi}$ over fixed Top-32 panels. The lower row shows the worst-block reference discrepancy $A_{\mathrm{ref},\chi}$ from Eq.~\eqref{eq:reference-adequacy} for SACE, EP-only, KL, and spectral selections. $Q_{\mathrm{highref},\chi}$ measures higher-resource MPS consistency rather than exact 50-qubit error; larger $A_{\mathrm{ref},\chi}$ indicates weaker convergence of the higher-resource reference.}
  \Description[50-qubit higher-resource validation]{The top row contains three scatter plots relating finite-bond convergence to higher-resource MPS consistency at bond dimensions 2, 4, and 8 for QFT and QAOA. The bottom row contains reference-adequacy box plots; the QAOA distributions lie substantially above the corresponding QFT distributions.}
  \label{fig:s11-reference-adequacy}
\end{figure}

Figure~\ref{fig:s11-reference-adequacy} reports the 50-qubit consistency and reference-convergence diagnostics defined in Eqs.~\eqref{eq:highref-validation} and~\eqref{eq:reference-adequacy}. The QFT references improve markedly with increasing $\chi$ and support the primary 50-qubit scale-up analysis. By contrast, the weighted-QAOA references remain weakly converged and are therefore interpreted only as a stress test. The corresponding medians and all weighted-QAOA-50 comparisons are retained in Table~\ref{tab:s5-50q-detailed}.

\FloatBarrier

\section{Detailed Large-Scale and Optimizer-Robustness Results}
\label{app:large-scale-tables}
Tables~\ref{tab:s4-24q-detailed} and~\ref{tab:s5-50q-detailed} separate large-scale results by circuit type, bond dimension, and baseline. The weighted-QAOA-50 rows are retained here for completeness as a reference-limited stress test: their large $A_{\mathrm{ref},\chi}$ values indicate that the higher-resource MPS reference is not sufficiently converged for these cases to serve as primary scale-up evidence. Table~\ref{tab:s6-optimizer-robustness} reports the combined optimizer study; search-overlap statistics are descriptive.

\begin{table}[!htbp]
\centering
\footnotesize
\setlength{\tabcolsep}{3pt}
\caption{24-qubit isolated-local exact-vs-MPS results for QFT and QAOA.}
\label{tab:s4-24q-detailed}
\textbf{(a) Outcomes and median errors}\par\smallskip
\begin{tabular*}{\linewidth}{@{\extracolsep{\fill}}lllllll@{}}
\toprule
Circuit & $\chi$ & Baseline & $n$ & W/L/T & SACE & Baseline \\
\midrule
QFT & 2 & EP-only & 10 & 10/0/0 & 0.3292 & 0.4376 \\
QFT & 2 & KL & 10 & 10/0/0 & 0.3292 & 0.4376 \\
QFT & 2 & Spectral & 10 & 10/0/0 & 0.3292 & 0.4376 \\
QFT & 4 & EP-only & 10 & 10/0/0 & 0.0597 & 0.2023 \\
QFT & 4 & KL & 10 & 10/0/0 & 0.0597 & 0.2023 \\
QFT & 4 & Spectral & 10 & 10/0/0 & 0.0597 & 0.2023 \\
QFT & 8 & EP-only & 10 & 10/0/0 & 0.0024 & 0.0297 \\
QFT & 8 & KL & 10 & 10/0/0 & 0.0024 & 0.0297 \\
QFT & 8 & Spectral & 10 & 10/0/0 & 0.0024 & 0.0297 \\
QAOA & 2 & EP-only & 10 & 7/2/1 & 0.9916 & 0.9939 \\
QAOA & 2 & KL & 10 & 4/5/1 & 0.9916 & 0.9870 \\
QAOA & 2 & Spectral & 10 & 3/6/1 & 0.9916 & 0.9915 \\
QAOA & 4 & EP-only & 10 & 6/4/0 & 0.9644 & 0.9687 \\
QAOA & 4 & KL & 10 & 5/5/0 & 0.9644 & 0.9619 \\
QAOA & 4 & Spectral & 10 & 5/5/0 & 0.9644 & 0.9574 \\
QAOA & 8 & EP-only & 10 & 8/2/0 & 0.7457 & 0.8873 \\
QAOA & 8 & KL & 10 & 9/1/0 & 0.7457 & 0.8323 \\
QAOA & 8 & Spectral & 10 & 8/1/1 & 0.7457 & 0.8148 \\
\bottomrule
\end{tabular*}

\medskip
\textbf{(b) Win rates and paired improvements}\par\smallskip
\begin{tabular*}{\linewidth}{@{\extracolsep{\fill}}llllll@{}}
\toprule
Circuit & $\chi$ & Baseline & Win rate & Median $\Delta Q$ & 95\% bootstrap interval \\
\midrule
QFT & 2 & EP-only & 1.000 & 0.0964 & [0.0766, 0.1524] \\
QFT & 2 & KL & 1.000 & 0.0964 & [0.0766, 0.1524] \\
QFT & 2 & Spectral & 1.000 & 0.0964 & [0.0766, 0.1524] \\
QFT & 4 & EP-only & 1.000 & 0.1319 & [0.0663, 0.1646] \\
QFT & 4 & KL & 1.000 & 0.1319 & [0.0663, 0.1646] \\
QFT & 4 & Spectral & 1.000 & 0.1319 & [0.0663, 0.1646] \\
QFT & 8 & EP-only & 1.000 & 0.0268 & [0.0101, 0.0343] \\
QFT & 8 & KL & 1.000 & 0.0268 & [0.0101, 0.0343] \\
QFT & 8 & Spectral & 1.000 & 0.0268 & [0.0101, 0.0343] \\
QAOA & 2 & EP-only & 0.700 & 0.0058 & [-0.0002, 0.0080] \\
QAOA & 2 & KL & 0.400 & -0.0006 & [-0.0071, 0.0019] \\
QAOA & 2 & Spectral & 0.300 & -0.0008 & [-0.0035, 0.0037] \\
QAOA & 4 & EP-only & 0.600 & 0.0161 & [-0.0072, 0.0307] \\
QAOA & 4 & KL & 0.500 & -0.0038 & [-0.0189, 0.0142] \\
QAOA & 4 & Spectral & 0.500 & -0.0005 & [-0.0212, 0.0152] \\
QAOA & 8 & EP-only & 0.800 & 0.1374 & [0.0130, 0.2203] \\
QAOA & 8 & KL & 0.900 & 0.1011 & [0.0454, 0.1580] \\
QAOA & 8 & Spectral & 0.800 & 0.0619 & [0.0103, 0.1147] \\
\bottomrule
\end{tabular*}

\begin{minipage}{0.98\textwidth}
\footnotesize $\Delta Q=Q_{\mathrm{baseline}}-Q_{\mathrm{SACE}}$; positive values favor SACE. The SACE and Baseline columns in panel (a) report median errors; panel (b) uses the same instance counts. Confidence intervals use an instance-cluster bootstrap. The reported quantity is isolated-local and is not reconstructed global-circuit fidelity.
\end{minipage}
\end{table}

\begin{table}[!htbp]
\centering
\footnotesize
\setlength{\tabcolsep}{3pt}
\caption{50-qubit higher-resource MPS consistency and reference-adequacy results. QFT provides the primary scale-up evidence; weighted-QAOA is retained as a reference-limited stress test.}
\label{tab:s5-50q-detailed}
\begin{tabular*}{\linewidth}{@{\extracolsep{\fill}}lllrlllll@{}}
\toprule
Circuit & $\chi$ & Baseline & $n$ & W/L/T & Win rate & \shortstack{Median\\$\Delta Q$} & \shortstack{Median\\$\rho_S$} & \shortstack{Median\\$A_{\mathrm{ref},\chi}$} \\
\midrule
QFT & 2 & EP-only & 10 & 6/3/1 & 0.600 & 0.0209 & 0.234 & 0.4274 \\
QFT & 2 & KL & 10 & 8/2/0 & 0.800 & 0.0525 & 0.234 & 0.4274 \\
QFT & 2 & Spectral & 10 & 8/2/0 & 0.800 & 0.0525 & 0.234 & 0.4274 \\
QFT & 4 & EP-only & 10 & 10/0/0 & 1.000 & 0.0882 & 0.822 & 0.1030 \\
QFT & 4 & KL & 10 & 10/0/0 & 1.000 & 0.0823 & 0.822 & 0.1030 \\
QFT & 4 & Spectral & 10 & 10/0/0 & 1.000 & 0.0823 & 0.822 & 0.1030 \\
QFT & 8 & EP-only & 10 & 9/1/0 & 0.900 & 0.0604 & 0.940 & 0.0167 \\
QFT & 8 & KL & 10 & 6/3/1 & 0.600 & 0.0142 & 0.940 & 0.0167 \\
QFT & 8 & Spectral & 10 & 6/3/1 & 0.600 & 0.0142 & 0.940 & 0.0167 \\
QAOA & 2 & EP-only & 9 & 6/3/0 & 0.667 & 0.0002 & 0.102 & 0.9751 \\
QAOA & 2 & KL & 9 & 4/5/0 & 0.444 & -0.0005 & 0.102 & 0.9751 \\
QAOA & 2 & Spectral & 9 & 5/4/0 & 0.556 & 0.0013 & 0.102 & 0.9751 \\
QAOA & 4 & EP-only & 9 & 7/2/0 & 0.778 & 0.0202 & 0.434 & 0.9025 \\
QAOA & 4 & KL & 9 & 3/6/0 & 0.333 & -0.0054 & 0.434 & 0.9025 \\
QAOA & 4 & Spectral & 9 & 6/3/0 & 0.667 & 0.0051 & 0.434 & 0.9025 \\
QAOA & 8 & EP-only & 9 & 9/0/0 & 1.000 & 0.0654 & 0.592 & 0.7503 \\
QAOA & 8 & KL & 9 & 7/2/0 & 0.778 & 0.0207 & 0.592 & 0.7503 \\
QAOA & 8 & Spectral & 9 & 7/2/0 & 0.778 & 0.0583 & 0.592 & 0.7503 \\
\bottomrule
\end{tabular*}
\begin{minipage}{\linewidth}
\footnotesize $Q_{\mathrm{highref},\chi}$ is higher-resource MPS consistency rather than exact 50-qubit error. The worst-block reference discrepancy $A_{\mathrm{ref},\chi}$ in Eq.~\eqref{eq:reference-adequacy} is reported for the SACE-selected partition. The weighted-QAOA-50 results cover nine instances and are retained only as exploratory stress-test evidence because the corresponding references remain weakly converged.
\end{minipage}
\end{table}

\begin{table}[!htbp]
\centering
\footnotesize
\setlength{\tabcolsep}{3pt}
\caption{Optimizer robustness and optimizer-pair stability.}
\label{tab:s6-optimizer-robustness}
\textbf{(a) Optimizer outcomes}\par\smallskip
\begin{tabular*}{\linewidth}{@{\extracolsep{\fill}}llllllll@{}}
\toprule
Optimizer & \shortstack{Own EP-only\\W/L/T} & \shortstack{Median\\$\Delta Q$} & vs KL & vs Spec. & \shortstack{Median\\$\rho_S$} & \shortstack{Oracle\\gap} & \shortstack{Archive\\size} \\
\midrule
GA & 23/6/1 & 0.0296 & 24/5/1 & 21/7/2 & 0.688 & 0.0098 & 1394--2987 \\
SA & 29/1/0 & 0.0716 & 28/2/0 & 28/2/0 & 0.744 & 0.0000 & 5917--5943 \\
DPSO & 24/5/1 & 0.0360 & 23/6/1 & 23/5/2 & 0.666 & 0.0022 & 1443--2635 \\
\bottomrule
\end{tabular*}
\medskip
\textbf{(b) Optimizer-pair stability}\par\smallskip
\begin{tabular*}{\linewidth}{@{\extracolsep{\fill}}lllll@{}}
\toprule
Pair & Median Jaccard & Equal selections & Median/max gap & Gap $>0.10$ \\
\midrule
GA-SA & 0.008 & 0 & 0.030/0.357 & 8 \\
GA-DPSO & 0.730 & 22 & 0.000/0.260 & 1 \\
SA-DPSO & 0.008 & 0 & 0.025/0.357 & 7 \\
\bottomrule
\end{tabular*}

\begin{minipage}{0.98\textwidth}
\footnotesize Optimizer rows use 30 cells each. Pair rows report the median Top-32 Jaccard overlap over ten instances, selected-partition equality over 30 cells, and absolute differences in isolated-local validation error $Q_{\mathrm{exact},\chi}$. Search diversity metrics are descriptive and did not affect selection.
\end{minipage}
\end{table}

\end{document}